\documentclass[runningheads]{llncs}

\usepackage[T1]{fontenc}
\usepackage{stmaryrd}
\usepackage{amssymb} 
\usepackage{amsfonts}
\usepackage{listings}
\usepackage{mathpartir}
\usepackage{tcolorbox}
\usepackage{colortbl}
\usepackage{tikz}
\usetikzlibrary{matrix,positioning}
\usepackage{bbding}
\usepackage{ebproof}

\usepackage{centernot}

\usepackage{graphics} % as of now, only for resizebox.
\usepackage{adjustbox} % as of now, only for resizebox.

\newcommand{\fakePar}[1]{\textit{#1:}}

\newcommand{\keydash}[2]{\key{#1}\texttt{-}\key{#2}}
\newcommand{\keydashThree}[3]{\key{#1}\texttt{-}\key{#2}\texttt{-}\key{#3}}

\newcommand{\myimplies}{\textit{ implies }}

\newcommand{\langLogic}{\mathbb{L}}
\newcommand{\assert}[3]{\{#1\}~#2~\{#3\}}
\newcommand{\assertOne}[1]{\{#1\}}

\newcommand{\topLevel}{\textit{constr}}
\newcommand{\getArgs}[1]{\textit{argsIdx}(#1)}

\newcommand{\expOne}{\textit{prg}_1}
\newcommand{\expTwo}{\textit{prg}_2}

\newcommand{\returnSymbol}{\hookleftarrow}

\newcommand{\ineffect}{\textsc{ineffectual}}

\newcommand{\selected}[1]{{selected}}

\newcommand{\print}[1]{\key{print} \app #1}

\newcommand{\reachableName}{\keydash{ctx}{compliant}}
\newcommand{\reachable}[1]{\reachableName(#1)}
\newcommand{\errHandlerName}{\keydash{handles}{error}}
\newcommand{\errHandler}[2]{\errHandlerName(#1,#2)}
\newcommand{\errHandlerShortName}{\textit{errhTry}}
\newcommand{\errHandlerShort}[2]{\errHandlerShortName}
\newcommand{\ctxPositions}[3]{\key{inductive}(#1,#2,#3)}
\newcommand{\effect}[1]{\key{effectful}}
\newcommand{\dupliEffectName}{\keydashThree{no}{dupli}{ef}}
\newcommand{\dupliEffect}[1]{\dupliEffectName(#1)}
\newcommand{\contravariantName}{\key{contravariant}}
\newcommand{\contravariant}[2]{\contravariantName(#1,#2)}
\newcommand{\variancePresName}{\keydash{contra}{resp}}
\newcommand{\variancePres}[2]{\variancePresName(#1,#2)}

\newcommand{\division}{\div}
\newcommand{\unitE}{\key{unit}}

\newcommand{\join}{\lor}

\newcommand{\quoting}[1]{``#1''}

\newcommand{\lna}{\textsc{Lang-n-Assert}}

\newcommand{\lnsql}{\textsc{Lang-Sql}}
\newcommand{\runningExample}{\lambda^{\division}_{\key{print}}}

\newcommand{\LangDef}{\mathcal{L}}

\newcommand{\langVar}[1]{{\textcolor{black}{#1}}}

\newcommand{\GammaForRule}[1]{\Gamma_{\textsf{rule}}}

\newcommand{\grammarDerivationG}{\Rightarrow_{G}^{*}}

\definecolor{magenta(dye)}{rgb}{0.79, 0.08, 0.48}
\definecolor{bondiblue}{rgb}{0.0, 0.58, 0.71}
\definecolor{navyblue}{rgb}{0.0, 0.0, 0.5}
\definecolor{lightskyblue}{rgb}{0.53, 0.81, 0.98}

\newcommand{\userLan}[1]{#1}

\newcommand*\colourcheck[1]{%
  \expandafter\newcommand\csname #1check\endcsname{\textcolor{#1}{\text{\ding{52}}}}%
}
\colourcheck{blue}
\colourcheck{green}
\colourcheck{red}

\newcolumntype{Y}{>{\raggedleft\arraybackslash}X}

\tcbset{tab1/.style={fonttitle=\bfseries\large,fontupper=\normalsize\sffamily,
colback=yellow!10!white,colframe=red!75!black,colbacktitle=red!40!white,
coltitle=black,center title,freelance,frame code={
\foreach \n in {north east,north west,south east,south west}
{\path [fill=red!75!black] (interior.\n) circle (3mm); };},}}

\tcbset{tab2/.style={enhanced,fonttitle=\bfseries,fontupper=\small\sffamily,
colback=yellow!10!white,colframe=red!50!black,colbacktitle=red!40!white,
coltitle=black,center title}}

\usepackage{mathtools}
\usepackage{relsize}

\newcommand{\ninference}[3]{\inferrule[(#1)]{#2}{#3}}

\newcommand{\ba}{\begin{array}}
\newcommand{\ea}{\end{array}}

\newenvironment{syntax}{\[\ba{l@{\;\;}lcl}}{\ea\]}

\definecolor{ShadowColor}{RGB}{30,150,190}

\makeatletter
\newcommand\Cshadowbox{\VerbBox\@Cshadowbox}
\def\@Cshadowbox#1{%
  \setbox\@fancybox\hbox{\fbox{#1}}%
  \leavevmode\vbox{%
    \offinterlineskip
    \dimen@=\shadowsize
    \advance\dimen@ .5\fboxrule
    \hbox{\copy\@fancybox\kern.5\fboxrule\lower\shadowsize\hbox{%
      \color{ShadowColor}\vrule \@height\ht\@fancybox \@depth\dp\@fancybox \@width\dimen@}}%
    \vskip\dimexpr-\dimen@+0.5\fboxrule\relax
    \moveright\shadowsize\vbox{%
      \color{ShadowColor}\hrule \@width\wd\@fancybox \@height\dimen@}}}
\makeatother

\newcommand{\typeOf}{\vdash}
\newcommand{\step}{\longrightarrow}

\newcommand{\unitT}{\key{Unit}}

\definecolor{lightblue}{rgb}{0.25,0.25,1}

\newcommand{\redd}[1]{\textcolor{red}{#1}}
\newcommand{\navy}[1]{\textcolor{lightblue}{#1}}
\newcommand{\magenta}[1]{\textcolor{magenta}{#1}}

\definecolor{lightgray}{gray}{0.9}
\definecolor{darkergrey}{rgb}{0.75, 0.75, 0.75}
\newcommand{\HI}[1]{\colorbox{darkergrey}{#1}}

\newcommand{\key}[1]{\ensuremath{\mathtt{#1}}}

\newcommand{\Int}{\key{Int}}

\newcommand{\Float}{\key{Float}}

\newcommand{\app}{\;}

\newcommand{\True}{\key{true}}

\definecolor{lightgray}{gray}{0.9}

\newcommand{\true}{{\key{true}}}

\newcommand{\try}[2]{\key{try} \app #1 \app \key{with} \app #2}

\newcommand{\Ldl}{\mathcal{L}}

\newcommand{\emptyLdl}[1]{\Ldl^{\epsilon}}

\usepackage{semantic}
\usepackage[breaklinks]{hyperref}
\usepackage[sort]{cite}

\title{From Program Logics to Language Logics}         %% [Short Title] is optional;
\titlerunning{From Program Logics to Language Logics}         %% [Short Title] is optional;

\author{Matteo Cimini}
\authorrunning{M. Cimini}

\institute{University of Massachusetts Lowell, Lowell MA 01854, USA
\\ \email{matteo\_cimini@uml.edu}
}

\begin{document}

\maketitle              

\begin{abstract}

Program logics are a powerful formal method in the context of program verification. 
Can we develop a counterpart of program logics in the context of language verification? 

This paper proposes \emph{language logics}, which allow for statements of the form $\assert{P}{\mathcal{X}}{Q}$ where $\mathcal{X}$, the subject of analysis, can be a language component such as a piece of grammar, a typing rule, a reduction rule or other parts of a language definition. 
To demonstrate our approach, we develop $\langLogic$, a language logic that can be used to analyze language definitions on various aspects of language design. 

We illustrate $\langLogic$ to the analysis of some selected aspects of a programming language. 
We have also implemented an automated prover for $\langLogic$, and we confirm that the tool repeats these analyses.

Ultimately, $\langLogic$ cannot verify languages. 
Nonetheless, we believe that this paper provides a strong first step towards adopting the methods of program logics for the analysis of languages. 
%\keywords{Language verification \and Language design \and Logic}
\end{abstract}
%qualitative aspects

% keywords?
%Language validation
%Formal methods for language verification
%Programming language design

\section{Introduction}\label{introduction}
 
%\redd{all-encompassing vs narrow, specific, localized, targeted?}

%\redd{global vs localized?}

Language verification is an important part of the development of programming languages. 
Once we have created a programming language, there are many questions that are interesting to investigate. 
%These questions vary greatly from concerning \emph{all-encompassing properties} of 
%the 
%%entire 
%language such as type soundness and relational parametricity to 
%concerning 
%These questions vary greatly, concerning both \emph{all-encompassing properties} of 
These questions vary greatly and they concern both \emph{all-encompassing properties} of 
a 
%entire 
language such as type soundness and relational parametricity 
as well as   
\emph{\selected{} aspects} of operators, grammar rules, and reduction rules, for example to determine whether %about whether 
the behaviour of our elimination forms is defined for all the expected values, whether we have defined all the necessary evaluation contexts, or whether contravariant arguments of type constructors are handled accordingly, to make a few examples. 
%Is the language type sound? is it strongly normalizing? is it free of data races? 

In the context of program verification, program logics stand out as a powerful formal method with decades of development and myriads of success stories. % \cite{Xu,Carbonneaux,Chen}. 
%Various program logics have been proposed over decades. 
Various program logics have been proposed in the literature. 
The seminal Floyd–Hoare logic has been applied to the verification of imperative programs \cite{Floyd1967Flowcharts,hoareProgramLogics,DijkstraProgramLogics}. 
Pointer-manipulating programs are better analyzed with separation logics \cite{ReynoldsSeparationL,reynolds2000intuitionistic,OHearnReynolds}, 
%\cite{ReynoldsSeparationL,reynolds2000intuitionistic,OHearnBI,OHearnReynolds}
while thread-based concurrent programs with concurrent separation logics \cite{OHearnConcurrent,BrookesCSL}. % SergeyCSL,AppelCSL,Haslbeck %Chargero
%Literature also offer logics that are specific to higher-order functional programs \cite{ArthurSLfunctional}, 
Literature also offers works on program logics that are specific to higher-order functional programs \cite{ArthurSLfunctional}, 
weak memory models \cite{SvendsenWeakMemoryLogic,VafeiadisRelaxedSeparation}, 
as well as many other domains \cite{Disantangle,Actris,Carbonneaux,Chen}. 
%and many other domains \cite{Disantangle,Actris,Carbonneaux,Chen}. 
%as well as to many other domains \cite{HOSL,Disantangle,Actris,Xu,Carbonneaux,Chen}.   shorter:  \cite{HOSL,Disantangle,Actris,Carbonneaux,Chen}. 
% JungIris, PottierVerification
%Program logics, too, have been applied to the all-encompassing correctness of programs as well as to smaller questions about parts of programs. 
%For example, program logics can analyze whether selected invariants hold before and after a loop that is in the middle of a large program. 

%Despite their power, there is no analogous in language verification for program logics. It has not been explored. 

% semi-automated tools based on separation logics: semi-automated tools based on Separation Logic, such as Infer [Calcagno et al., 2015], VeriFast [Philippaerts et al., 2014], or Viper [Müller et al., 2016].

Our question: \emph{Can we develop a counterpart of program logics for the verification of languages?} 

\paragraph{Language Logics.}

We propose \emph{language logics}. 
In language logics, the subject of analysis is a language definition rather than a program. 
Statements in language logics have the form $\assert{P}{\mathcal{X}}{Q}$ where $\mathcal{X}$ can be the entire language at hand or some of its components such as a piece of grammar, a typing rule, a reduction rule or other parts of the language definition. 
Analogously to program logics, $P$ is a precondition and $Q$ is a postcondition. 
To make an example, given an inference rule $r$, 
%(say, a typing rule), 
$\assert{P}{r}{Q}$ can be read \quoting{when $P$ holds, $Q$ holds after having added the inference rule $r$ to the language definition}.

To demonstrate our approach, we have developed $\langLogic$, a language logic that can be used to analyze language definitions on various aspects of programming languages. 
%that is capable of reasoning about various aspects of programming languages. 
%(These are \emph{local aspects}, following the distinction made at the beginning of this section.)  
Assertions of $\langLogic$ ($P$ and $Q$ above) can be built with formulae that are domain-specific to the context of language design. 
The aim of these formulae is to reason about \emph{\selected{} aspects}, following the distinction made at the beginning of this section. 
For example, $\langLogic$ can express an assertion $\contravariant{c}{\{i_1, \ldots, i_n\}}$ that means that the arguments of the type constructor $c$ at positions $i_1$, $\ldots$, $i_n$ are contravariant. Also, $\langLogic$ can express the assertion $\effect{i}$ that means that the language is effectful, i.e., operations can modify a state. Similarly, the assertion $\reachable{rn}$ means that if the reduction rule with name \textit{rn} needs some expressions to be values in order to fire, then the corresponding evaluation contexts are in place for those arguments to be evaluated. 
%(We provide only a few examples here in the introduction. Section \ref{logic} will provide the full range of formulae of $\langLogic$.) 
%(Section \ref{logic} will provide the full range of formulae of $\langLogic$.)
Section \ref{logic} will provide the full range of formulae of $\langLogic$. 
%The aim of these formulae is to reason about \emph{\selected{} aspects}, following the distinction made at the beginning of this section. 
%Their aim is to reason about \emph{\selected{} aspects}, following the distinction made at the beginning of this section. 
%They address \emph{\selected{} aspects}, following the distinction made at the beginning of this section. 
%These model \emph{local aspects}, following the distinction made at the beginning of this section. 

We define the proof rules of $\langLogic$ in the style of program logics. These proof rules derive statements $\assert{P}{\mathcal{X}}{Q}$ where assertions $P$ and $Q$ involve the formulae that we have described. 
%The proof rules of $\langLogic$ \redd{are based on} detecting common syntactic patterns and may not guarantee a property if the language also has ill-behaved inference rules. 
The proof rules of $\langLogic$ detect common syntactic patterns for deriving assertions. 
As we point out in \S \ref{limitations}, they may not \emph{guarantee} a property. 
%As we point out in Section \ref{limitations}, they may not \emph{guarantee} a property. 
%The implication is that they may not \emph{guarantee} a property, as we shall discuss in Section \ref{limitations}.

%\redd{we say upfront that this is the first exploration and we did not come up with a full verification method.}

%We provide the syntax, semantics and proof rules of $\langLogic$. 
%
%\[
%%\ninference{t-app-bad}
%\inference
%	{
%	 \Gamma \typeOf \app e_1 : T_1\to T_2 \\
%	 \Gamma \typeOf \app e_2 : T_3 & \HI{$T_1 <: T_3$}  	
%	} 
%	{  \Gamma \typeOf \app e_1\app e_2 : T_2}
%\quad
%\assert{\contravariant{\to}{\{1\}}}{\textsc{(t-app-bad)}}{\redd{no proof rule applicable}}	
%\]
%
%

\paragraph{Evaluation: The Language Logic $\langLogic$ at Work.}

%To demonstrate our language logic $\langLogic$, we embark on a journey towards debugging the definition of a faulty language. % with (floating point) divisions, functions, error handlers, subtyping, and print-effects that modify a string buffer. %that we intend to design. 
To demonstrate our language logic, we embark on a journey towards debugging the definition of a faulty language. % with (floating point) divisions, functions, error handlers, subtyping, and print-effects that modify a string buffer. %that we intend to design. 
This language has a few issues, for example it duplicates effects due to a call-by-name strategy 
and does not take into account that the domain of function types is contravariant. 
%and its evaluator may fail the overall computation at the encounter of an error without passing the error to the error handler. % the chance to handle the error differently. 
Each time that we detect an issue, we show that $\langLogic$ cannot, indeed, derive the corresponding assertion. % that we intended to achieve. 
%We show, then, that after we modify the language definition to fix the issue we now can provide such proof derivation in $\langLogic$. % of the property that we sought for. 
We show, then, that after we modify the language and fix the issue we now can provide such proof derivation in $\langLogic$. % of the property that we sought for. 

We have implemented an automated prover for $\langLogic$ called $\lna$ \cite{lna}. % as an assertion prover called $\lna$. 
%The tool takes in input the elements of a statement: a precondition assertion, a language definition, and a postcondition. 
%Language definitions are a textual representation of operational semantics, an example is in Appendix \ref{lan}. 
Given a statement $\assert{P}{\LangDef}{Q}$, the tool provides a proof derivation for it or fails, 
if a derivation is not found. 
We confirm that $\lna$ replicates the debugging journey of the faulty language, 
failing to derive sought for assertions and succeeding upon fixing the issues. %after progressively fixing the language. 
%We confirm that $\lna$ replicates the successes and failures of the above debugging journey of the faulty language, 
%failing to derive the sought for assertion at first and then succeeding after progressively fixing the language. 

%We acknowledge that the verification of languages is not available to language logics yet. %
We acknowledge that the verification of languages is not available to language logics yet, as we lack a soundness theorem and do not capture \emph{all-encompassing properties}. 
We offer a discussion of these challenges in Section \ref{limitations}. 
%We offer a discussion of these challenges in \S \ref{limitations}. 
%We discuss these challenges in \S \ref{limitations}. 
Nonetheless, we believe that this paper provides a strong first step towards adopting the methods of program logics for the analysis of languages. 
%in the context of language analysis. 

%This is the first paper on language logics and we leave several issues unexplored (Section \ref{limitations}). 
%%We offer a comparison between program logics and language logics (Section \ref{comparison}), 
%%and we discuss the limitations of language logics (Section \ref{limitations}). 
%For example, it is unclear how to apply language logics to \emph{global properties} such as type soundness, and there are difficulties in achieving a soundness and completeness theorem for $\langLogic$. 
%%This is the first paper exploring a counterpart of program logics for language definitions. 
%%Part of our future work is to fill this gap. 

The paper is organized as follows. 
Section \ref{operational} reviews the elements of operational semantics.
Section \ref{logic} provides the syntax and proof rules of $\langLogic$. 
Section \ref{examples} applies $\langLogic$ to the analysis of our running example. 
Section \ref{comparison} offers a comparison between program logics and language logics. 
Section \ref{limitations} discusses the limitations of language logics. 
Section \ref{related} discusses related work 
and Section \ref{conclusion} concludes the paper.

\section{Operational Semantics (Review)}
\label{operational}

%We focus on languages defined with operational semantics. 
%To review, we show the language definition of our running example in Fig. \ref{fig:language}. 
Fig. \ref{fig:language} shows the language definition of our running example $\runningExample$. 
%To review, we show the language definition of our running example in Fig. \ref{fig:language}. 
This is a $\lambda$-calculus with integers, floating points, subtyping, a simple \key{try} error handler, and a \key{print} operation that adds strings into a buffer. 
%(This could model a simplified version of printing on screen that does not provide the more precise possibility to pixels and their values.) 
%We call this language $\runningExample$. 

A language has a grammar which consists of a series of \emph{grammar rules}, each of which defines a \emph{syntactic category}, such as \textsf{Type} and \textsf{Expression}. 
Each syntactic category has a metavariable, such as $T$ and $e$, and \emph{grammar productions}, such as $\Int$, $\Float$, and $T\to  T$ of \textsf{Type}. 
%\emph{Terms}, ranged over by $t$, are the elements that can be derived by a grammar (they possibly contain metavariables). 
%%Terms can contain metavariables (such as grammar productions) or not (such as $\Bool \to \Bool$, for example). 
%%Terms can contain metavariables, as grammar productions do. 
%Terms can use unary binding $(X)t$ \cite{Cheney:2005}. %, which means that the variable $X$ is bound in $t$. 
%%, and capture-avoiding substitution $t[t/X]$. 
%Terms can use unary binding $(X)t$ \cite{Cheney:2005}, denoting that the variable $X$ is bound in $t$, and capture-avoiding substitution $t[t/X]$. 
%As $\lnp$ needs to access terms uniformly, terms will be handled in abstract syntax tree, 
%for example $(\times \app T_1 \app T_2)$ rather than $T_1 \times T_2$.  
%i.e., a top-level constructor applied to arguments. 
A language also has inference rules that define relations such as a typing, a subtyping, and a reduction relation. 
Each inference rule has a series of formulae called \emph{premises} and one formula called \emph{conclusion}. 
For example, $\Gamma \typeOf \app e_1 :\Float$ and $\Gamma \typeOf \app e_2 : \Float$ are premises of rule \textsc{[t-div]}, and $\Gamma \typeOf \app e_1\division e_2 : \Float$ is its conclusion. 
%$\runningExample$ defines several relations. 
Inference rules whose conclusion can derive a $\typeOf$-formula are called \emph{typing rules}, 
those that derive a $<:$-formula are called \emph{subtyping rules}, and 
those that derive a $\step$-formula are called \emph{reduction rules}. 
%The typing and subtyping relations of $\runningExample$ have a standard shape ($\Gamma \typeOf \app e : T$, with a standard typing environment $\Gamma$, and $T<:T$, respectively). 
%The typing and subtyping relations of $\runningExample$ have a standard shape. 
$\runningExample$ has standard typing and subtyping relations. % of  have a standard shape. 
% ($\Gamma \typeOf \app e : T$, with a standard typing environment $\Gamma$, and $T<:T$, respectively). 
The reduction relation of $\runningExample$ is of the form $e,s \step e',s'$ where $e$ is the expression to be evaluated and $s$ is the state of the computation. The state is a string buffer. %that contains the current content . 
The evaluation reduces $e$ to $e'$ and may lead to a modified state $s'$. 
The only operation that modifies the state is $\key{print}$. %, which appends a string to the current string buffer. 
As typical, we use the \key{Unit} type for a side-effect. 
%As typical, we use the \key{Unit} type for a side-effect of the like. 

\emph{Evaluation contexts} declare which arguments of an expression constructor can be evaluated, and also in which order they are to be evaluated. 
\emph{Error contexts} define in which contexts we are allowed to detect the occurrence of an error and fail the overall computation. 
This is realized with rule \textsc{[err-ctx]}. 
In $\runningExample$, the error \key{error} is generated after a division by $0$. 
%Programmers can catch the occurrence of \key{error} with the \key{try} error handler. 
% and specify an alternative computation other than simply fail the overall computation. 

%\paragraph{Issues with $\runningExample$} 
\fakePar{Issues with $\runningExample$} 
%The language definition of $\runningExample$ in Fig. \ref{fig:language} contains a few issues. 
The language definition of $\runningExample$ contains a few issues. 
%(None of these issues, nor the fixes adopted in Section \ref{examples}, are a novelty of this paper.) 
(None of these issues, nor their fixes in Section \ref{examples}, are a novelty of this paper.) 
\textit{Issue 1:} \textsc{[cbn-beta]} adopts a call-by-name strategy in the presence of effects. 
%This may lead to the unexpected duplication of \key{print}-effects. 
This may lead to the unpredictable duplication of \key{print}-effects. % (especially if we do not know how many times a function makes use of the formal argument at run-time). 
\textit{Issue 2:} The error context $\try{F}{e}$ entails that the evaluator may \quoting{steal} the error from the error handler and terminate the computation rather than letting \key{try} handle the error. % occurrence. 
\textit{Issue 3:} \textsc{[t-app-bad]} mistakes the direction of the subtyping relation between the domain of the function and the type of the argument. 
%This means that, for example, we will not be able to pass an integer to a function that requests a floating point. 
This means, for example, that we cannot pass an integer to a function that requests $\Float$. 
% a floating point. 

It would be desirable to reason about these issues using proof derivations in the style of program logics.
%It would be desirable to validate that these issues do not occur with proof derivations in the style of program logics.

\begin{figure}[tbp]
\small
%\textit{Warning: There are a few issues in this language definition.}\\
%\textit{Warning: There are a few issues highlighted in this language definition, which we discuss in Section \ref{operational}.}\\
\textit{Disclaimer: }
\indent \textit{There are a few issues, highlighted, which we discuss in Section \ref{operational}.}\\
%\textit{There are a few issues highlighted in this language definition, which we discuss in Section \ref{operational}.}\\

%\begin{itemize} 
%\item \textsc{(t-app)} mistakes the direction of the subtyping relation between the domain of the function and the type of the argument. 
%\item A call-by-name strategy may lead to the unexpected duplication of \key{print}-effects. 
%\item Error contexts may \quoting{steal} the error from \key{try} rather than letting \key{try} handle the error. 
%\end{itemize} 

$n\in\mathbb{N}, f\in\mathbb{R},s \in \textsc{String}$
\begin{syntax}
  \textsf{Type} & T & ::= &\Int \mid \Float \mid T\to  T \mid \unitT\\ %\mid \RefType \app T\\
     \textsf{Expression} & e & ::= &   n\mid f \mid {e} \division {e}
%     \\
%     &&&
     \mid x  \mid \lambda x:T.e\mid (e\app e) \\ % \mid \compose{e}{e}{e}\\
     &&& \mid \unitE  \mid \print{s} \mid e;e
       \\ 
 &&& 
  \mid \key{error} \mid \try{e}{e}  \\
  \textsf{Value} & v & ::= & 
  n 
  \mid f
  %\true \mid \false  
\mid\lambda x:T.e 
\mid \unitE
\\
   \textsf{Error} & \mathit{er} & ::= &\key{error}  \\
  \textsf{EvalCtx} & E & ::=  &  
  \Box 
   \mid E\division e\mid v\division E 
   \mid (E\app e) % cbn \mid (v\app E) 
   \mid E ; e 
 %\mid \key{raise}\app E
  \mid \try{E}{e} \\
% &&& \mid \myLet{x}{E}{e} \\
  \textsf{ErrorCtx} & F & ::=  &  
  \Box 
   \mid F\division e\mid v\division F 
   \mid (F\app e) % cbn \mid (v\app F) 
   \mid F ; e 
 %\mid \key{raise}\app F 
 \mid \HI{$\try{F}{e}$} \\
%  \textsf{TypeEnv} & \Sigma & ::=  & ....... \\
%  \textsf{TypeEnv} & \Gamma & ::=  & \emptyset \mid \Gamma,x:T\\
\end{syntax}
\small
Type System  \hfill  \fbox{$ \Gamma \vdash e : T$}
\begin{gather*}
% Bool
%\ninference{t-int}{}
%\ninference{t-float}{}
{ \Gamma\typeOf n:\Int}\qquad
{ \Gamma\typeOf f : \Float}\qquad
%\ninference{t-div}
\inferrule[\textsc{[t-div]}]
	{
	 \Gamma \typeOf \app e_1 : \Float 
	 %\\
	 \quad
	 \Gamma \typeOf \app e_2 : \Float 
	} 
	{  \Gamma \typeOf \app e_1\division e_2 : \Float}
%\ninference{t-var}{}
\\[1ex]
{ \Gamma,x:T\typeOf\kern -1pt x:T}
%\ninference{t-label}
%\ninference{t-true}{}
%{\Gamma \typeOf \app \true:\Bool}
%\qquad 
%\ninference{t-false}{}
%{\Gamma \typeOf \app \false:\Bool}	
%\\[0.5ex]
%\inference
%	{\Gamma \typeOf \app e_1 : \Bool &
%	\Gamma \typeOf \app e_2 : T &
%	\Gamma \typeOf \app e_3 : T 
%	} 
%	{ \Gamma \typeOf \app\ifExp{e_1}{e_2}{e_3} : T}  %\,\,\textsc{(t-if)}
%\\[0.5ex]
%\ninference{t-abs}
\qquad 
\inference
	{ \Gamma,x:T_1 \typeOf \app e : T_2} 
	{  \Gamma \typeOf \app \lambda x:T_1. e:  T_1\to T_2}
%\\[1ex] 
\qquad
%\ninference{t-app-bad}
%\inference
\inferrule[\textsc{[t-app-bad]}]
	{
	 \Gamma \typeOf \app e_1 : T_1\to T_2 
	 %\\
	 \quad
	 \Gamma \typeOf \app e_2 : T_3 
	 \\\\ 
	 \HI{$T_1 <: T_3$}  	
	} 
	{  \Gamma \typeOf \app e_1\app e_2 : T_2}
	\qquad
%\ninference{t-error}
%{ \Gamma \typeOf \app e:\HI{$T$}}
%	{\Sigma \mid \Gamma \typeOf \app \myerror{e} : \HI{$T$}}
\\[1ex]
\begin{array}{l}
\Gamma\typeOf \unitE:\unitT\\[0.5ex]
\Gamma\typeOf \print{s}:\unitT\\[0.5ex]
\Gamma\typeOf \key{error}: T
\end{array}
~~~
%\qquad
\inference
	{
	 \Gamma \typeOf \app e_1 : T_1 &
	 \Gamma \typeOf \app e_2 : T_2 
	} 
	{  \Gamma \typeOf \app e_1; e_2 : T_2}
~~~
%\qquad
%\ninference{algo-t-try}
\inference
{\Gamma \typeOf \app e_1 :  T_1 &
	\Gamma\typeOf e_2 : T_2 \\
	%\Bool \subA \app T_3 \\ 
	T_1 \join T_2 = T_j
		}
	{ \Gamma \typeOf \app \try{e_1}{e_2} :  T_j}
%\ninference{t-try}
%{\Gamma \typeOf \app e_1 :  T \\\\
%	\Gamma\typeOf e_2 : \Int \to T 	}
%	{ \Gamma \typeOf \app \try{e_1}{e_2} :  T}
\end{gather*}
Subtyping \hfill  \fbox{$T<:T$}
\begin{gather*}
%\ninference{s-base}{}
%%\inference
%\begin{array}{l}
\Int <: \Float%\\[0.5ex]
%T <: T
%\end{array}
%\Float <: \Float\\\\ 
%\Int <: \Float
%\,\,\textsc{(t-head)}
%${\begin{array}
%\Int <: \Float
%\end{array}
%}$
\qquad
\begin{array}{l}
\Int <: \Int\\[0.5ex]
\Float <: \Float\\[0.5ex]
\unitT <: \unitT
\end{array}
\qquad
%\inference
%{T_1 <: T_2 & T_2 <: T_3}
%{T_1 <: T_3}
%\qquad
%\ninference{s-arrow}
\inferrule[\textsc{[s-arrow]}]
%\inference
	{T_1' <: T_1 
	%\\
	\quad 
	T_2 <: T_2' }
{ T_1 \to T_2 <: T_1' \to T_2'}%\,\,\textsc{(t-head)}
%\qquad
%  \inferrule[Foo]{A \quad B}{C}
%\\[0.5ex]
\end{gather*}
Reduction Semantics  \hfill \fbox{$e,s \step e,s$}
\begin{align*}
%%	\ifExp{\true}{e_1}{e_2} & \step  e_1 %\label{r-if-true}\tagsc{r-if-true}
%%	\\[-5pt]
%%	\\
%%	\ifExp{\false}{e_1}{e_2} &  \step  e_2  % 	\label{r-if-false}\tagsc{r-if-false}
%%	\\[-5pt]
%%	\\
	f_1 \division f_2,  s & \step  f_3, s ~~~(f_2 \not=0)  & \textsc{[div]} %\label{beta}\tagsc{div}
	\\[-3pt]
	f_1 \division 0,  s & \step  \key{error}, s \label{beta} &  %\textsc{[div-err]} %~~~(f_2 \not=0)\label{beta}\tagsc{div}
	\\%[-5pt]
	(\lambda x:T. e_1)\app \HI{$e_2$}, s& \step  e_1[\HI{$e_2$}/x], s  & \textsc{[cbn-beta]}
%	(\lambda x:T. e)\app \HI{$e'$}, s& \step  e[\HI{$e'$}/x], s  \qquad\textsc{(cbn-beta)} %\qquad (\beta\app \mathit{rule})%\label{beta}\tagsc{beta}
	\\[-3pt]
	\print{s_2}, s_1 & \step   \unitE, s_1 + \text{\quoting{$\returnSymbol$}} + s_2    	& \textsc{[print]} 
	\\[-3pt]
	v;e, s & \step   e,s    	& %\textsc{[seq]} 
	\\[-3pt]
	\try{v}{e}, s & \step  v, s   & %\textsc{[try]} 
	\\[-3pt]
	\try{\key{error}}{e}, s  & \step  e, s& \textsc{[err]} 
%	\try{\key{raise}\app v}{e}  & \step  (e \app v)% 	\label{beta}\tagsc{r-let} 
\end{align*}
$$
%\ninference{ctx}
\inference
	{e,s\step e',s'}
	{E[e],s \step E[e'],s'}%\,\,\textsc{(ctx)}
\qquad 
%\ninference{err-ctx}
\inferrule[\textsc{[err-ctx]}]
{}
{F[\mathit{er}],s \step \mathit{er},s} %\,\,\textsc{(err-ctx)}\\[3ex]
$$\\
%\textit{where $f_3$ is the result of the division between $f_1$ and $f_2$, and $+$ is string concatanation}
\textit{where $f_3$ is the division between $f_1$ and $f_2$, $+$ is string concatanation, and $\returnSymbol$ is the newline return symbol}
\caption{Language definition of $\runningExample$  
%(typing and subtyping). We use notation $T \subA\app T$ for algorithmic subtyping rather than the more standard $\typeOfA T <: T$  \cite{tapl}. 
}
\label{fig:language}
\end{figure}

\subsection{A Syntax for Language Definitions}\label{languageDefinitions}

Program logics such as Floyd–Hoare logic work with statements $\assert{P}{c}{Q}$ where $c$ is a command with a formal syntax. %so that formal proof rules can be devised. 
Language logics analyze a language $\LangDef$ with statements $\assert{P}{\LangDef}{Q}$. 
Analogously to program logics, $\LangDef$ must be accommodated with a formal syntax. 
%Analogously to program logics, language definitions $\LangDef$ must be accommodated with a formal syntax. 
%
%We adopt the syntax for language definitions of prior work \cite{multilanguage}, which is simply a grammar for operational semantics definitions. % that we recalled above. 
%
%The syntax is the following.\\
We adopt the following syntax for language definitions from prior work \cite{multilanguage}, which is simply a grammar for operational semantics definitions. % that we recalled above. 
%
%The syntax is the following.\\

%The syntax for (programmer-defined) language definitions is the following. %below in Figure \ref{fig:syntax:loml}. 

{\footnotesize
\noindent 
\quad
$
  \userLan{cname} \in \textsc{CatName}, ~\userLan{\langVar{X}} \in \textsc{MetaVar}, ~  \userLan{pn} \in \textsc{PredName},~ \userLan{rn} \in \textsc{RuleName}\\
%  \indent 
  \quad 
  \userLan{c},op \in \textsc{ConstructorName}  
  ~(\text{We use \emph{op} when we know it to be an operator}.)
%  ~(\text{we use \emph{op} in the next sections})
%  \indent 
%  \userLan{pn} \in \textsc{PredName}, \userLan{rn} \in \textsc{RuleName}
$
\begin{syntax}
   \text{\sf Language} & \userLan{\LangDef} & ::= & \userLan{(G,I)} \\
   \text{\sf Grammar} & \userLan{G} & ::= & \userLan{ g_1 \app \cdots \app g_n } \\
   \text{\sf {Grammar Rule}} & \userLan{g} & ::= & \userLan{cname \app \langVar{X} ::= t_1\app  \mid \cdots \app \mid \app  t_n}\\
   \text{\sf Inference System} & \userLan{I} & ::= & \userLan{ r_1 \app \cdots \app r_n }\\
   \text{\sf Rule} & \userLan{r} & ::= & rn: \userLan{\inference{f_1 \app \cdots \app f_n }{f}}\\
   \text{\sf Formula} & \userLan{f} & ::= & \userLan{(pn \app t_1 \cdots\app  t_n)}  \\
   \text{\sf Term} & \userLan{t} & ::= &  \userLan{\langVar{X}} \mid \userLan{(c \app t_1 \cdots\app  t_n)} \mid \userLan{(\langVar{X})t}  \mid \userLan{t[t/\langVar{X}]} %\\ % \mid \userLan{(\langVar{X})t  \mid t[t/\langVar{X}] \\
%   \text{\sf List of Rules} & \userLan{R} & ::= & \nil \mid \consLNC{r}{R}\\
%   \text{\sf List of Formula} & lf & ::= & \nil \mid \consLNC{f}{lf}\\
%   \text{\sf List of Terms} & lt & ::= & \nil \mid \consLNC{t}{lt}
\end{syntax}
}
%\caption{Language Definitions in $\calcName$.}
%\label{fig:syntaxLanguage}
%\end{figure*}
%
\textsc{CatName} contains syntactic category names. % such as \textsf{Expression} and \textsf{Type}.  
\textsc{MetaVar} contains metavariables. % such as $E$ and $T$,  
\textsc{ConstructorName} contains constructor names. % such as \key{\Int}, \key{\to}, and $\lambda$ (they do not have to necessarily be (string) names).   
\textsc{PredName} contains predicate names (names of relations), and  % such as $\typeOf$, $\step$, and $<:$. 
\textsc{RuleName} contains names of inference rules such as \textsc{[cbn-beta]}. 
Terms are accommodated with a uniform syntax in abstract syntax style (top-level name applied to arguments). 
Formulae, as well, are in abstract syntax. 
For readability, however, we will use familiar syntax such as $e_1\step e_2$, $\Gamma \typeOf e : T$, $(e_1\app e_2)$, $T_1\to T_2$, and so on, in our examples. 
Terms can also use unary binding $\userLan{(\langVar{X})t}$ \cite{Cheney:2005} and capture-avoiding substitution $\userLan{t[t/\langVar{X}]}$. 
% Below does not say that you use familiar syntax in examples. 
%\textsc{RuleName} contains names of inference rules such as \textsc{[cbn-beta]} and \textsc{[err-ctx]}. 
%Terms are accommodated with a uniform syntax in abstract syntax style with a top-level constructor name applied to arguments. Terms can also use unary binding $\userLan{(\langVar{X})t}$ \cite{Cheney:2005} and capture-avoiding substitution $\userLan{t[t/\langVar{X}]}$. 
%Formulae, as well, are in abstract syntax style. 

Prior works \cite{multilanguage,lnc1,lns} have shown examples of operational semantics definitions in this syntax. 
%Our language 
$\runningExample$, too, can be accommodated as a language $\LangDef$. 
%We show only a few parts: grammar of types, the typing rule for divisions, and \textsc{(cbn-beta)}. 
%
%\begin{gather*}
%  \textit{Type} \app T  ::= \textit{int} \mid \textit{float} \mid (\to \app T\app T) \mid \textit{unitT}
%  \\[1ex] 
%%  \textsc{(cbn-beta)}: (\step \app ((\lambda x:T. e_1)\app e_2) \app s \app  (e_1[e_2/x]) \app  s) 
%%
%%\ninference{t-app-bad}
%\textsc{(t-div)}: \inference
%	{
%	  (\typeOf \app \Gamma \app e_1 \app \textit{float}) &
%	  (\typeOf \app \Gamma \app e_2 \app \textit{float}) 
%	} 
%	{   (\typeOf \app \Gamma \app (\division \app e_1\app e_2) \app \textit{float})  }
%	\\[1ex]
%%
%\textsc{(cbn-beta)}: (\step \app ((\lambda x:T. e_1)\app e_2) \app s \app  (e_1[e_2/x]) \app  s) 
%%	(\lambda x:T. e_1)\app e_2, s \step  e_1[e_2/x], s  
%\end{gather*}

%Some languages produce effects on a state. 
%We assume that the syntactic categories of terms that \emph{do not} produce effects are declared in $\ineffect$. 
%%We assume that the syntactic categories of terms that \emph{do not} produce effects are known, and we assume that their metavariables can be found in a set that we call $\ineffect$. 
%For example, $\{v,er\} \subseteq \ineffect$ in $\runningExample$ because values and errors cannot produce effects. 
%(Our logic will make use of $\ineffect$ to detect, for example, whether a reduction rule is not susceptible to effect duplication for handling ineffectual terms.) 

\section{$\langLogic$: A Language Logic for Analyzing Languages}\label{logic}

%In this section, we develop the language logic $\langLogic$. 
%Section \ref{syntax} provides the syntax of $\langLogic$, and Section \ref{rules} provides its proof rules. 

\subsection{Syntax of $\langLogic$}\label{syntax}

%Fig. \ref{fig:syntax} defines the syntax of our language logic $\langLogic$. 
The following is the syntax of our language logic $\langLogic$. 
The design idea behind the assertions of $\langLogic$ is that they state a specific aspect that is of interest in the context of language design. 
%about the type constructors, operations, reduction rules, typing rules, and so on, of the language at hand. %as well as about the overall language. 
We have selected a handful of formulae. % to demonstrate our approach. 
%We have selected a handful of formulae that address various programming languages aspects. 
By no means they are all that it would be interesting to detect of a language. 
{\small
%\noindent 
%$
%  \userLan{cname} \in \textsc{CatName}, \userLan{\langVar{X}} \in \textsc{Meta-Var}, 
%  \userLan{opname} \in \textsc{OpName}, \userLan{pn} \in \textsc{PredName}
%$
\begin{syntax}
   \text{\sf Assertion} & P,Q & ::= & 
\ctxPositions{\userLan{X}}{c}{\{n_1, \ldots, n_k\}}
%\\&&&
\mid       \reachable{rn}
\\ &&&
      \mid \errHandler{op}{n}
%\\ &&&
\mid
\effect{\userLan{X}}
\mid
\dupliEffect{op}
\\ &&&
\mid 
\contravariant{c}{n}
\mid \variancePres{rn}{c}
\\ 
%&&&
%\mid
%\extrusion{op}\\
    &&& \mid \True \mid  P \land Q  \mid \lnot P \\  %  \mid P \Rightarrow Q and and not is functionally complete. 
   \begin{array}{c}
%   \text{\sf Annotated} \\ \textsf{Lang/Grammar/Rule} 
   \text{\sf Annotated} \\ \textsf{Language}  \\ \textsf{Component} 
   \end{array}& ~~ & ::= & \assert{P}{\LangDef}{Q} \mid \assert{P}{\userLan{G}}{Q} \mid  \assert{P}{\userLan{I}}{Q}\mid  \assert{P}{g}{Q} \mid  \assert{P}{r}{Q}
\end{syntax}
}

The assertion $\ctxPositions{\userLan{X}}{c}{\{n_1, \ldots, n_k\}}$ holds whenever $c$ is a top-level constructor of a grammar production of the category with metavariable $X$ and its arguments at positions $n_1$, $\ldots$, and $n_k$ are \emph{inductive} in that they are $X$ also. 
%In its general form, $\ctxPositions{\userLan{X}}{op}{n_1, \ldots, n_k}$ holds whenever the arguments of $op$ at positions $n_1$, $\ldots$, and $n_k$ are all $X$s, as well. 
To make an example, $\ctxPositions{\userLan{T}}{\to}{\{1,2\}}$ holds because the two arguments of the function type in $\textsf{Type} \app T ::= \ldots \mid T\to  T$ are inductive. 
%We will use this assertion in Section \ref{examples} to derive the evaluated arguments of an operator. For example, $\ctxPositions{\userLan{E}}{\division}{\{1,2\}}$ means that an evaluation context is declared for both arguments of division, that is, the two arguments of division are evaluated. 

The assertion $\reachable{rn}$ means that if the reduction rule with name $rn$ needs some arguments to be values (or errors) in order to fire, then the corresponding evaluation contexts are in place for those arguments. % to evaluate. 
%To see why we use the name \quoting{reachable}, let us consider the reduction rule $\textsc{[div]}$, which requires $f_1 \division f_2$ to fire. 
%To see why we use the name \quoting{reachable}, let us consider the reduction rule $\textsc{[div]}$, which requires $f_1 \division f_2$ to fire. 
%To make an example, $\textsc{(div)}$ requires its source to be of the form $f_1 \division f_2$. 
%To see why $\reachableName$ is an interesting formula, let us consider rule $\textsc{[div]}$, which requires $f_1 \division f_2$ to fire. 
To see what $\reachableName$ tells us, let us consider rule $\textsc{[div]}$, which requires $f_1 \division f_2$ for \emph{values} $f_1$ and $f_2$ to fire. 
Given a division $(e_1\division e_2)$, the existence of evaluation contexts $E\division e$ and $v\division E$ means $\reachable{\textsc{[div]}}$ and 
%therefore 
%also means 
that 
$e_1$ and $e_2$ may have a chance to become $f_1$ and $f_2$ for $\textsc{[div]}$. 
%$e_1$ and $e_2$ may become $f_1$ and $f_2$ so that $\textsc{[div]}$ may have a chance to fire. 
%$e_1$ and $e_2$ may have a chance to become $f_1$ and $f_2$ so that $\textsc{[div]}$ may have a chance to fire. 
%A rule is not guaranteed to apply. For example, the two reduction rules of the if-then-else statement (for \textit{true} and \textit{false}) are \quoting{ctx-compliant} but after the guard is evaluated only one of them fires. 
%, so that $\textsc{[div]}$ has a chance to take over the computation from there. Therefore, $\reachable{\textsc{[div]}}$ holds. 

%Notice that the rule is not guaranteed to fire, for example even if 

The assertion $\errHandler{op}{n}$ holds whenever a reduction rule for \textit{op} exists that is \quoting{ctx-compliant}, handles an error as $n$-th argument of \textit{op}, and error contexts are unable to detect the error at that position. 
%The assertion $\errHandler{op}{n}$ holds whenever $op$ is such that if an error occurs as its $n$-th argument then there is a reduction rule of $op$ that fires in lieu of failing the overall computation. 
%is an error handler and appropriately handle errors that occur at its $n$-th argument. 
%That is, there is a rule that describe the behavior of $op$ and that rule only 

The assertion $\effect{\userLan{X}}$ holds whenever the language has a state and has reductions that can modify the state. 
%For example, $\effect{s}$ holds in the context of $\runningExample$. 
%The assertion 
$\dupliEffect{rn}$ holds whenever the step of the reduction rule $rn$ does not duplicate arguments that may produce effects. 
%the effects that are prescribed to occur once at the evaluation of one of its arguments. 

The assertion $\contravariant{c}{\{n_1, \ldots, n_k\}}$ holds whenever the arguments of the type constructor $c$ at positions $n_1$, $\ldots$, $n_k$ are contravariant. For example, $\contravariant{\to}{\{1\}}$ holds for the function type. 
%
%The assertion $\variancePres{rn}{c}$ 
$\variancePres{rn}{c}$ 
holds for the typing rule $rn$ whenever the types in the premises of $rn$ that appear as contravariant arguments of $c$ are not used at the left of a subtyping formula. 
%holds for the typing rule $rn$ whenever the types in the premises of $rn$ that appear as contravariant arguments of $c$ are not used at the lefthand side of a subtyping formula. 

%$\langLogic$ also has conjunction and negation (which is a complete set of connectives.) 
%$\langLogic$ has conjunction and negation (which form a complete set of connectives.) 

An \emph{annotated language} is a language with a pre- and postcondition: $\assert{P}{\LangDef}{Q}$ means that \quoting{when $P$ holds, $Q$ holds after having analyzed the language $\LangDef$}. 
Similarly, we have an annotated grammar, grammar rule, inference system, and inference rule. 
%Similarly, we have an \emph{annotated grammar} $\assert{P}{G}{Q}$, \emph{annotated grammar rule} $\assert{P}{g}{Q}$, \emph{annotated inference system} $\assert{P}{I}{Q}$, and \emph{annotated inference rule} $\assert{P}{r}{Q}$. 
The meaning of these is analogous to that of annotated languages. 
For example, $\assert{P}{r}{Q}$ means that \quoting{when $P$ holds, $Q$ holds after having added the inference rule $r$}. 
We write $\{\}$ in lieu of $\assertOne{\true}$. 
The typical use of our logic is to start analyzing $\LangDef$ from $\{\}$ and derive $\assert{}{\LangDef}{Q}$, for an assertion $Q$.

\subsection{Proof Rules of $\langLogic$}\label{rules}

Fig. \ref{fig:rules} and \ref{fig:rulesRest} define the proof rules of $\langLogic$. 
Fig. \ref{fig:rules} shows the proof rules that govern the traversal of languages and their components as well as the composing of assertions. 
%handle languages and their components while 
Fig. \ref{fig:rulesRest} shows the proof rules that analyze single grammar rules $g$ and single inference rules $r$ in order to derive the assertions of the previous section. 
%Fig. \ref{fig:rulesRest} shows the proof rules that analyze single grammar rules $s$ and single inference rules $r$ in order to derive assertions that are specific to the context of language design, such as those that we have seen in the previous section. 

%\noindent 
We first discuss the proof rules of Fig. \ref{fig:rules}. 
%The design principle around them is that they analyze a language by reading its components one after another. 
The design principle that they follow is that they analyze a language by reading its components one after another. 
%as they are encountered in the language definition. %That is, we start from the first 
%Each time, the assertions that are derived are \quoting{passed} to the analysis of the rest of the language definition. 
Each time, the assertions that are derived are \quoting{passed} to the analysis of the rest. % of the language. % definition. 
%Each time, the assertions so derived are \quoting{passed} to the analysis of the rest of the language. % definition. 
%Each time, the assertions that are derived are \quoting{passed} to the analysis of the rest of the language. % definition. 
%Each time, the assertions that are derived from a language component are \quoting{passed} to the analysis of the rest of the language definition. 

\begin{figure}[tbp]
\small
%Declarative Type System  \hfill  \fbox{$\Gamma \vdash e : T$}
\begin{gather*}
% Bool
%\mathcal{X}\in\{\LangDef, G, I, g, r\}
\ninference{lang}
	{
	\assert{P}{G}{Q} \\\\
	\assert{Q}{I}{R}	
	} 
	{\assert{P}{(G,I)}{R}}% \,\,\textsc{(t-app)}
%\\[1ex]
\qquad 
\ninference{grammar}
	{
	\assert{P_0}{g_1}{P_1} \\\\
	\assert{P_1}{g_2}{P_2} \\\\
	\ldots \\\\
	\assert{P_{n-1}}{g_n}{P_n} 
	} 
	{  \assert{P_0}{g_1 \app \cdots \app g_n}{P_n}}% \,\,\textsc{(t-app)}
%\qquad 
%\ninference{perm-grammar}
%	{  \assert{P}{s_1 \app \cdots s_{i+1} \app  s_i  \app \cdots \app s_n}{Q}}% \,\,\textsc{(t-app)}
%	{  \assert{P}{s_1 \app \cdots s_i \app s_{i+1} \app \cdots \app s_n}{Q}}% \,\,\textsc{(t-app)}
%\\[1ex]
\qquad
\ninference{inf}
	{
	\assert{P_0}{r_1}{P_1} \\\\
	\assert{P_1}{r_2}{P_2} \\\\
	\ldots \\\\
	\assert{P_{n-1}}{r_n}{P_n} 
	} 
	{  \assert{P_0}{r_1 \app \cdots \app r_n}{r_n}}% \,\,\textsc{(t-app)}
\qquad 
\\[1ex]
\ninference{perm-g}
	{  \pi \textit{ is a permutation of } g_1 \app \cdots \app g_n \\\\
	\assert{P}{\pi}{Q}}% \,\,\textsc{(t-app)}
	{  \assert{P}{g_1 \app \cdots \app g_n}{Q}}% \,\,\textsc{(t-app)}
\qquad
\ninference{perm-r}
	{  \pi \textit{ is a permutation of } r_1 \app \cdots \app r_n \\\\
	\assert{P}{\pi}{Q}}% \,\,\textsc{(t-app)}
	{  \assert{P}{r_1 \app \cdots \app r_n}{Q}}% \,\,\textsc{(t-app)}
\qquad
\\[1ex]
\begin{array}{l}
\ninference{$\mathcal{X}$-neutral}
%{\textit{Admissible rule}}
{}
{\assert{P}{\mathcal{X}}{P}}\\[2ex]
\ninference{iterate}
	{
%	P \Rightarrow P' \\
	\assert{P}{\mathcal{X}}{Q} \\
	\assert{Q}{\mathcal{X}}{R} 
%	Q' \Rightarrow Q \\
	} 
	{\assert{P}{\mathcal{X}}{R}}% \,\,\textsc{(t-app)}
\end{array}
\qquad
\ninference{consequence}
	{
	P \Rightarrow P' \\
	\assert{P'}{\mathcal{X}}{Q'} \\
	Q' \Rightarrow Q \\
	} 
	{\assert{P}{\mathcal{X}}{Q}}% \,\,\textsc{(t-app)}
\qquad 
%\\[1ex]
%\mathcal{X}\in\{\LangDef, G, I, s, r\}
\end{gather*}
\caption{Main proof rules of $\langLogic$. We have $\mathcal{X}\in\{\LangDef, G, I, g, r\}$.}
%\caption{Proof rules of $\langLogic$. We have $\mathcal{X}\in\{\LangDef, G, I, g, r\}$.}
\label{fig:rules}
\end{figure}

Proof rule \textsc{(lang)} analyzes the grammar of the language and, starting from the assertions so derived, analyzes the inference system. 
Proof rule \textsc{(grammar)} analyzes the grammar rules, one by one, in the order they are encountered. Each time, the assertions derived from a grammar rule are used as preconditions in the analysis of the next grammar rule. Proof rule \textsc{(inf)} is analogous to \textsc{(grammar)} and analyzes the inference rules in the order they are encountered. 
Proof rule \textsc{(perm-g)} allows to analyze the grammar rules in any order. Similarly, rule \textsc{(perm-r)} allows to analyze the inference rules in any order. 
%Combined with \textsc{(grammar)}, rule \textsc{(perm-g)} allows to analyze the grammar rules of the language in any order. Similarly, rule \textsc{(perm-r)} allows to analyze the inference rules in any order. 

Proof rule \textsc{($\mathcal{X}$-neutral)} propagates the precondition as postcondition. % any new assertions from a language component. 
%We use this rule to simply traverse a language component and propagate the assertions previously derived to the next steps. 
%For example, on our way to build a proof derivation, we may have derived assertion $P$ when we get to analyze the grammar rule, say, $\textsf{Error}\app er := \key{error}$. Suppose that there is no information that we need from this grammar rule. 
%We then simply can use \textsc{($\mathcal{X}$-neutral)} to derive $\assert{P}{\textsf{Error}\app er := \key{error}}{P}$ and pass $P$ to the analysis of the rest of the grammar rules and inference system. 
Rule \textsc{(iterate)} analyzes a language component to derive $Q$. 
%Proof rule \textsc{(iterate)} analyzes a language component to derive $Q$. 
Then, it analyzes again the same language component using $Q$ as precondition. 
%Then, the rule analyzes again the same language component using $Q$ as precondition. 
%As \textsc{(iterate)} can be used recursively, it can be used to keep the focus of the analysis on the same language component and apply multiple rounds of proof rules to derive multiple assertions from it. 
The last proof rule of Fig. \ref{fig:rules} corresponds to the standard \textsc{(consequence)} rule of program logics. 
This rule allows for the strengthening of preconditions and the weakening of postconditions. 
%We will discuss \textsc{($\mathcal{X}$-neutral)}, \textsc{(iterate)}, and \textsc{(consequence)} further after we address Fig. \ref{fig:rulesRest}. 

%We now discuss the proof rules of Fig. \ref{fig:rulesRest}. 
%The design principle around these rules is the following: 
%The design principle of the rules of Fig. \ref{fig:rulesRest} is the following: 
The design principle of the proof rules of Fig. \ref{fig:rulesRest} is: 
%The design principle of the proof rules of Fig. \ref{fig:rulesRest} is the following: 
For each of the assertions in the grammar of $P$, we provide one or more proof rules that can derive that assertion. 
Such derivation is based on detecting common syntactic patterns. 
%(Here, we describe the rules and do not repeat the meaning of assertions.) 
%(Below, we do not repeat the meaning of formulae, we only described the rules.) 

\begin{figure}[tbp]
\small
%Declarative Type System  \hfill  \fbox{$\Gamma \vdash e : T$}
%\textit{In this figure, we assume that proof rules for $\assert{P}{r}{Q}$ can refer to the grammar $G$ of the language is accessible.} \\
\textit{We assume that proof rules of $\assert{P}{r}{Q}$ can use the grammar $G$ of the language.} \\
\textit{In this fig., we use symbols $s$ for the terms of reduction formulae that form the state.}
%\textit{In this fig., we use symbols $s$ for the terms that form the state in a reduction formula.}
\begin{gather*}
\ninference{inductive}    
	{
	I \subseteq \{1 \ldots n\}\\\\
	\forall i\in I, t_i.\topLevel = c \\ \forall i\in (\{1,\ldots, n\} - I), t_i.\topLevel \not= c\\\\
	I' = \bigcup_{i\in I} t_i.\getArgs{X} 
	} 
	{\assert{P}{\userLan{cname \app \langVar{X} ::= t_1\app  \mid \cdots \app \mid \app  t_n}}{P\land \ctxPositions{\userLan{X}}{c}{I'}}}%\,\,\textsc{(t-head)}
% below with j \in I instead. 
%\ninference{inductive}    
%	{
%	I \subseteq \{1 \ldots n\}\\\\
%	\forall j\in I, t_j.\topLevel = c \\ \forall j\in (\{1,\ldots, n\} - I), t_j.\topLevel \not= c\\\\
%	I' = \bigcup_{j\in I} t_j.\getArgs{X} 
%	} 
%	{\assert{P}{\userLan{cname \app \langVar{X} ::= t_1\app  \mid \cdots \app \mid \app  t_n}}{P\land \ctxPositions{\userLan{X}}{c}{I'}}}%\,\,\textsc{(t-head)}
\\[3ex]
\ninference{ctx-compliant}
{ \ctxPositions{\userLan{E}}{op}{I} \in P \\\\
\forall i, 1 \leq i \leq n, (err \grammarDerivationG t_i \lor v \grammarDerivationG t_i) \myimplies i \in I
%\\
%\widetilde{s} \textit{ are distinct variables}
} % \{n_1, \ldots, n_k\}}
{\assert{P}{rn: (op \app t_1, \ldots, t_n), \widetilde{s} \step t, \widetilde{s'}}{P \land \reachable{rn}}}
%{\assert{P}{rn: (op \app t_1, \ldots, t_n), s_1, \ldots, s_m \step t, s'_1, \ldots, s'_m}{P \land \reachable{rn}}}
\\[3ex]
\ninference{error-handler}
{ \reachable{rn} \in P \\
\ctxPositions{\userLan{F}}{op}{I} \in P \\\\
err \grammarDerivationG t_i 
%\\\\
\\
%I_{err} \subseteq I_{ctx} \\ 
i \not\in I
}
{\assert{P}{rn: (op \app t_1, \ldots, t_n), \widetilde{s} \step t,  \widetilde{s'}}{P \land \errHandler{op}{i}}}
\\[3ex]
\ninference{effectful}
{ s_i \not= s_i'
}
{\assert{P}{rn : (op \app \widetilde{t}), s_1, \ldots, s_m \step t, s'_1, \ldots, s'_m}{P \land \effect{i}}}
\\[3ex]
%\ninference{avoid-dupli-effect}
%{
%\effect{i} \in P \\
%\forall i, e \grammarDerivationG t_i \Rightarrow 
%\lnot(G \typeOf t = E[t_i, t_i, \ldots] \lor G \typeOf t = E[t'[t_i/x]])}
%{\assert{P}{rn : (op \app t_1, \ldots, t_n), s_1, \ldots, s_m \step t, s'_1, \ldots, s'_m}{P \land \dupliEffect{rn}}}
%\\[3ex]
\ninference{effectual-args}
{
\effect{i} \in P \\\\
\forall t\in\widetilde{t}, (\nexists X\in \ineffect, X \grammarDerivationG t) 
\myimplies {~~~~~~~~~~~~~~~~~~}
\\\\
~~~~~~~~~~~~~~~~~~~~~~~~~~~~(\lnot(t' \textit{is of the form } C[t, t, \ldots]){~}
\land {~} \lnot(t' \textit{is of the form } C[t''[t/x]]))}
{\assert{P}{rn : (op \app \widetilde{t}\,), \widetilde{s} \step t', \widetilde{s'}}{P \land \dupliEffect{rn}}}
\\[3ex]
%\ninference{contravariant}
%{I=\bigcup_{T_i' <: T_i \in \widetilde{f}} \{i\}}
%{\assert{P}{rn : \inference{\widetilde{f}}{(c \app T_1 \app \ldots \app T_n) <: (c \app T_1' \app \ldots \app T_n') }}{P \land \contravariant{c}{I}}}
\ninference{contravariant}
{}
{\assert{P}{rn : \inference{\widetilde{f}}{(c \app T_1 \app \ldots \app T_n) <: (c \app T_1' \app \ldots \app T_n') }}{P \land \contravariant{c}{\bigcup_{(T_i' <: T_i) \in \widetilde{f}} \{i\}}}}
\\[3ex]
\ninference{contra-respecting}
{
\contravariant{c}{I} \in P\\\\
%\forall i,\app  f_i = t_1' \typeOf t_2' : (c \ap ) \Rightarrow t_3'.args(T') \subseteq I}
%\forall i,\app f_i = T <: T' \Rightarrow \forall j,\app  f_j = t_1' \typeOf t_2' : t_3' \Rightarrow t_3'.args(T') \subseteq I}
%\forall j,\app  f_j = t'' \typeOf t''' : (c \app t_1 \cdots \app t_m) \Rightarrow \forall i \in I, t_i}
\forall f\in \widetilde{f}, \app  f = t'' \typeOf t''' : (c \app t_1 \cdots \app t_m) \myimplies \forall i \in I, \lnot\exists f'\in \widetilde{f}, f' = t_i <: t_i'}
{\assert{P}{rn : \inference{\widetilde{f}}{\Gamma \typeOf t : t'}}{P \land \variancePres{rn}{c}}}
%{\assert{P}{rn : \inference{f_1 \ldots f_m}{\Gamma \typeOf (op \app t_1 \app \ldots \app t_n) : t}}{P \land \variancePres{rn}{c}}}
\end{gather*}
%\caption{Proof rules of $\langLogic$ for $\assert{P}{cname \app \langVar{X} ::= t}{Q}$ and $\assert{P}{r}{Q}$}
\caption{
%Proof rules of $\langLogic$ for $\assert{P}{g}{Q}$ and $\assert{P}{r}{Q}$. 
Proof rules for $\assert{P}{g}{Q}$ and $\assert{P}{r}{Q}$. 
Notation $\,\widetilde{\cdot}\,$ denotes finite sequences. 
%In this figure,  notation $\,\widetilde{\cdot}\,$ denotes finite sequences. 
}
\label{fig:rulesRest}
\end{figure}

%$\ctxPositions{\userLan{X}}{c}{\{n_1, \ldots, n_k\}}$ where 
Proof rule \textsc{(inductive)} analyzes a grammar rule and derives an assertion $\ctxPositions{\userLan{X}}{c}{I'}$ where $I'$ is a set of indices (of arguments' positions). 
In this rule, $t.\topLevel$ returns the top-level constructor of $t$, e.g., $(\to \,T \app T).\topLevel = \,\to$. 
Also, $t.\getArgs{X}$ returns the positions of the arguments of $t$ that are equal to $X$, e.g., $(\to \,T \app T).\getArgs{T} = \{1,2\}$. 
%The productions of the grammar rule are the terms $t_1$, $\ldots$, $t_n$. 
The terms $t_1$, $\ldots$, $t_n$ are the grammar productions. 
The rule focuses on those terms whose top-level constructor is $c$ (with $ t_i.\topLevel = c$). 
The premise $\forall i\in (\{1,\ldots, n\} - I), t_i.\topLevel \not= c$ makes sure that we select all of them. % within the grammar rule. 
For each, we extract the position of their arguments that are $X$ with $ t_i.\getArgs{X}$. 
% below with j instead. 
%The rule focuses on those terms whose top-level constructor is $c$ (with $ t_j.\topLevel = c$). 
%The premise $\forall j\in (\{1,\ldots, n\} - I), t_j.\topLevel \not= c$ makes sure that we select all of them. % within the grammar rule. 
%For each, we extract the position of their arguments that are $X$ with $ t_j.\getArgs{X}$. 
These positions are combined together in $I'$.

Proof rule \textsc{(ctx-compliant)} analyzes a reduction rule of \textit{op} with name \textit{rn}.  
This rule derives an assertion $\reachable{rn}$. 
%The precondition $\ctxPositions{\userLan{E}}{op}{I_{ctx}}$ says that the arguments of $op$ at positions $I_{ctx}$ are subject to an evaluation context. 
Assertion $\ctxPositions{\userLan{E}}{op}{I}$ is the precondition and says that the arguments of $op$ at positions $I$ are subject to an evaluation context. 
For example, $\ctxPositions{\userLan{E}}{\division}{\{1,2\}}$. 
% that we know which arguments of \textit{op} are subject to an evaluation context (with $\ctxPositions{\userLan{E}}{op}{I_{ctx}} \in P$). 
The rule checks whether any argument of \textit{op} is required be a value or an error for the rule to fire.  
This is checked with the standard grammar derivation $\grammarDerivationG$ to see if those arguments are derived from the metavariable of values $v$ or errors $err$. 
%For each of those arguments, we check that an evaluation context has been declared for them. 
For each, we check that an evaluation context has been declared for them. 
%To ensure that no other conditions prevent the rule from firing, \textit{rn} has no premises (only conclusion) and the state consists of distinct variables (which unifies with any state). 
%Notice that this does not ensure that a reduction can be proved with \textit{rn}. 
Notice that this does not mean that a reduction is sure to occur using \textit{rn}. 
For example, rule \textit{rn} may require the state to unify with a state that does not occur at runtime. 
%never occurs at runtime. 

Proof rule \textsc{(error-handler)} analyzes a reduction rule of \textit{op} with name \textit{rn} and derives $\errHandler{op}{i}$. 
%The precondition is that we have previously derived that the rule is reachable (with $\reachable{rn} \in P)$ and that we know which error contexts apply to the arguments of \textit{op} (with $\ctxPositions{\userLan{F}}{op}{I} \in P$). 
The preconditions are $\reachable{rn}$ and $\ctxPositions{\userLan{F}}{op}{I}$. 
The latter informs about the error contexts for \textit{op}. 
%the assertion that informs about the error contexts for \textit{op} ($\ctxPositions{\userLan{F}}{op}{I} \in P$). 
If the rule requires an argument to be an error to fire, then we check that such argument is \emph{not} subject to an error context (with $i \not\in I$). 

Proof rule \textsc{(effectful)} analyzes a reduction rule 
%that defines the behavior of an operator \textit{op} 
and derives an assertion $\effect{}$ whenever  one of the terms in the state is modified after the step. 
%The rule checks that one of the terms in the state is modified after the step. 

Proof rule \textsc{(effectual-args)} analyzes a reduction rule of \textit{op} with name \textit{rn} and derives an assertion $\dupliEffect{rn}$. %, but they do so based on two different cases. 
The precondition is that the language is effectful (with $\effect{} \in P$). 
%We assume that the set $\ineffect$ contains the metavariables of categories that are known not to produce effects. 
We assume that the language designer knows which syntactic categories do not produce effects. 
We assume that the set $\ineffect$ contains the metavariables of these categories. % that are known not to produce effects. 
For $\runningExample$, $\ineffect = \{v,err\}$.  
%To do that, t
The rule focuses on the arguments that cannot be derived by any of the categories of $\ineffect$ (with $\nexists X\in \ineffect, X \grammarDerivationG t$). 
(These arguments may produce effects which may be duplicated.) 
For each, we check that the target of the step does not replicate it nor use it in a substitution operation. (The rule makes use of context-like notation: $C[t, t, \ldots]$ for a term that contains $t$ two or more times, and $C[t''[t/x]]$ for a term that contains a substitution that involves $t$.) 
%We could instead use a set for effectual categories such as $\textsc{effectual} = \{e\}$ but we would need to handle the fact that values are also expressions. 

Proof rule \textsc{(contravariant)} analyzes the subtyping rule of a type constructor $c$ and derives an assertion $\contravariant{c}{\{i_1, \ldots, i_n\}}$. 
%For all premises I contains the index $i$. 
The rule computes the indices $i_1$, $\ldots$, $i_n$ from premises of the rule of form $T_i' <: T_i$. 
%the set $I$ by all the indices i such that a premise %\bigcup_{(T_i' <: T_i) \in \widetilde{f}} \{i\}}

Proof rule \textsc{(contra-respecting)} analyzes a typing rule with name \textit{rn} 
% that defines the behavior of an operator \textit{op}. 
%This rule derives an assertion $\variancePres{rn}{c}$. 
and derives an assertion $\variancePres{rn}{c}$. 
%The precondition is that we know which arguments of a type constructor \textit{c} are contravariant (with $\contravariant{c}{I} \in P$). 
%The precondition is that we know which arguments of a type constructor \textit{c} are contravariant (with $\contravariant{c}{I} \in P$). 
The precondition is the assertion that informs about the arguments of the type constructor \textit{c} that are contravariant (with $\contravariant{c}{I} \in P$). 
The rule detects those typing premises whose output type is built with \textit{c} as top-level constructor, 
i.e., formulae of the form $t'' \typeOf t''' : (c \app t_1 \cdots \app t_m)$. For each of the arguments $t_1$, $\cdots$, $t_m$ that $I$ labels as contravariant, say $t_i$,  we check that the rule does not contain any premise where $t_i$ 
appears at the left of a subtyping formula 
%appears on the lefthand side of a subtyping relation 
(with $\lnot\exists f'\in \widetilde{f}, f' = t_i <: t_i'$). 
%Notice that we have not used negated assertions $\lnot P$. %we have not used it. 
%Appendix \ref{negation} provides an example of a possible proof rule that involves negation. 
%\redd{$\lna$ is publicly available and contains all our tests \cite{lna}.}

\paragraph{Some Remarks on Proof Derivations.}
%\fakePar{Some Remarks on Proof Derivations}
The rules in Fig. \ref{fig:rulesRest} are the base case of our proof system. 
Each of them \emph{adds} a new assertion. %into the postcondition. 
However, \textsc{(consequence)} can be used to \quoting{forget} assertions due to $P \land Q \Rightarrow Q$. 
(Appendix \ref{forget} shows this application.) %of \textsc{(consequence)}.)
%(Appendix \ref{forget} shows this application of \textsc{(consequence)}.)
%However, some assertions may be needed up to some point in a proof and never be used again. 
%Rule \textsc{(consequence)} can \quoting{forget} assertions as follows: 
%%
%\[
%\ninference{consequence}
%	{
%	P \Rightarrow P \\
%	\assert{P}{\mathcal{X}}{P \land Q} \\
%	P \land Q \Rightarrow Q \\
%	} 
%	{\assert{P}{\mathcal{X}}{Q}}% \,\,\textsc{(t-app)}
%\]
%
%where the postcondition has been weakened and $P$ has been discarded. 
%When some of the 
Section \ref{examples} illustrates our language logic and when some proof derivations forget assertions it will be understood that we applied \textsc{(consequence)}. 
%Some proof derivations in Section \ref{examples} forget assertions and it will be understood that we applied \textsc{(consequence)}. 
% in the way shown above. 

%Our proof rules may derive \emph{one} assertion from an inference or grammar rule. 
Proof rules of Fig. \ref{fig:rulesRest} derive \emph{one} assertion from an inference or grammar rule. 
To derive multiple assertions from the same inference or grammar rule, \textsc{(iterate)} can be used recursively. 
(Page \pageref{iterate} shows an example on deriving two assertions.) 
%(An example is at page \pageref{iterate}.) 
%\textsc{(iterate)} can be used to keep analyzing the same inference or grammar rule to derive multiple assertions from it. 

There may be inference rules $r$ for which no proof rule applies and no statement $\assert{P}{r}{Q}$ can be derived. 
This is a problem with rule \textsc{(inf)}, as it would get stuck while traversing all rules. 
Rule \textsc{($\mathcal{X}$-neutral)} ensures that we can at least propagate the assertion so far derived. 
% to the rest of the analysis. 
Notice that \textsc{(consequence)} does not simulate \textsc{($\mathcal{X}$-neutral)} by weakening when its central premise cannot be derived. 
%\redd{$\lna$ is publicly available and contains all our tests \cite{lna}.}

The order in which rules are analyzed may matter. For example, subtyping rules must be analyzed before typing rules in order to derive a $\contravariantName$ assertion for $\variancePresName$. 
This motivates the permutation rules of Fig. \ref{fig:rules}. 
%This motivates \textsc{(perm-r)} in our proof system for the reordering of rules (and \textsc{(perm-g)} for analogous reasons). 

\section{Evaluation: The Language Logic $\langLogic$ at Work}\label{examples}

%  \textsf{Type} & T & ::= & \TopT  \mid \Int \mid \app T\to  T \mid \RefType \app T\\
%     \textsf{Expression} & e & ::= &   \topE\mid n \\
%%     &&&\mid \key{if}\app e\app \key{then}\app e\app \key{else}\app e \\
%     &&&\mid x  \mid \lambda x:T.e\mid (e\app e) 
%  \\ 
% &&& 
%  \mid \key{raise}\app e \mid \try{e}{e}  \\

%In this section, we embark on a journey to test and debug $\runningExample$. 
In this section, we embark on a journey to analyze $\runningExample$. 
Each time that we encounter an issue, we show that $\langLogic$ cannot indeed derive its corresponding assertion. 
We show then that $\langLogic$ can make such derivation after fixing the issue.

\paragraph{Issue 1a: Duplicating Effects.}
%\fakePar{Issue 1a: Duplicating Effects}
%
%The following program makes use of a \key{print}-effect before passing the number $2$ to the function, but the effect is being duplicated. 
The following program makes use of a \key{print}-effect before passing $2$ to the function, but the effect is being duplicated. 
%The following program makes use of a \key{print}-effect before passing the number $2$ to the function, but the effect is being duplicated. 
%
\begin{gather*}
\expOne = ((\lambda x: \Float. \app x \division x)  \app\app (\print \text{\quoting{\textit{Performing a parameter passing}}} ; 2)), \, \text{\quoting{\,\!\,\!}} \\%[2ex]
%\expOne, \text{\quoting{}} \step^{*} \app 
\expOne \step^{*} \app 
2 \division 2, 
\begin{array}{l}
%\quad 
\text{``}\textit{Performing a parameter passing} \\
 %\quad\quad\quad\quad
 ~\textit{Performing a parameter passing}\,\text{''}
\end{array}
\end{gather*}
%

%As \textsc{[print]} does \emph{not} print a string twice on its own, the problem could be that some operator duplicates an argument that may produce effects. 
The problem is that function application duplicates an argument that may produce effects. 
Let us try to derive \dupliEffect{\textsc{[cbn-beta]}}. We do so starting from the precondition $\effect{}$. 
(We assume it for now but we show its derivation shortly.) 
%(We show a derivation of $\effect{}$ after we fix this issue.) 
When we try to use \textsc{(effectual-args)}, however, there are premises that we cannot satisfy. 
Below, the facts that do not satisfy the rule are in red color. 
(Recall that $\ineffect = \{v,er\}$.) 
\begin{gather*}
%\ninference{ineffectual-args}
%{
%%\effect{i} \in P \\
%v\in \ineffect 
%%\\ 
%~\land~
%v \grammarDerivationG (\lambda x:T. e_1) \textit{ (* first arg of $(\lambda x:T. e_1)\app e_2$ *)}
%%\\ 
%\\\\
%\redd{v \not\grammarDerivationG e_2 ~\land~ er \not\grammarDerivationG e_2} \textit{ (* second arg of $(\lambda x:T. e_1)\app e_2$ *)}
%%\redd{v \not\Rightarrow e_2} \textit{ (* second arg of $(\lambda x:T. e_1)\app e_2$ *)}
%%\redd{~\land~ v \not\Rightarrow e_2}
%}
%{\assert{\effect{i}}{\textsc{[cbn-beta]}:(\lambda x:T. e_1)\app e_2, s \step  e_1[e_2/x], s}
%{ \app \ldots \app }
%}
%\\[1ex]
\ninference{effectual-args}
{
v\in \ineffect 
~\land~
v \grammarDerivationG (\lambda x:T. e_1) \textit{ (* \quoting{implies} vacuously satisfied *)} 
\\\\
(\nexists X\in \ineffect, X \grammarDerivationG e_2)
%\Rightarrow 
%\lnot(G \typeOf e_1[e_2/x] = E[e_2, e_2, \ldots] \lor G \typeOf t' = E[t''[t/x]])}
%\redd{~\land~e_1[e_2/x] = E[t'[e_2/x]]}
\redd{~\land~e_1[e_2/x] \textit{ is of the form } C[t'[e_2/x]]}
} 
{\assert{\effect{i}}{\textsc{[cbn-beta]}:(\lambda x:T. e_1)\app e_2, s \step  e_1[e_2/x], s}
{ \app \ldots \app }
%{\ldots \app N/A \app \ldots}
}
\end{gather*}

We cannot apply \textsc{(effectual-args)} because $e_2$ cannot be classified as ineffectual, 
and the target of the step does use $e_2$ in a substitution (which may replicate $e_2$). 
%We can debug our language and replace rule \textsc{(cbn-beta)} with its call-by-value counterpart, which is known not to suffer from effects' duplication: $\textsc{(beta)}: (\lambda x:T. e)\app \HI{$v$}, s \step  e_1[\HI{$v$}/x], s$.  
We can debug our language by replacing \textsc{[cbn-beta]} with (call-by-value)\footnote{An alternative way to fix this issue could be to retain \textsc{[cbn-beta]}, make the functional language pure, and include monads, for example. However, we did not go that way.} $\textsc{[beta]}: (\lambda x:T. e)\app \HI{$v$}, s \step  e[\HI{$v$}/x], s$.  
Below is a derivation of $\dupliEffect{\textsc{[beta]}}$ for the whole language.

%\textit{Convention:} Formulae that do not indicate the name of a proof rule are derived with \textsc{($\mathcal{X}$-neutral)} unless otherwise specified. % proved shortly after. 

%\textit{Convention:} Formulae that do not indicate the name of a proof rule are derived with \textsc{($\mathcal{X}$-neutral)} unless they have a color, which means that their proof derivation is given shortly afterwards using that same color. %their respective color. 

\noindent \textit{Convention:} Formulae without a proof rule name are derived with \textsc{($\mathcal{X}$-neutral)}. Colored formulae are derived afterwards using the same color. 
% unless otherwise specified. % proved shortly after. 

%
%
\[
%\hspace{-1em}
{\small
\begin{prooftree}
%        \hypo{A}
%        \infer1{B}
%        \hypo{C}
        \hypo{\assert{}{G}{}}
%        \hypo{\assert{}{\textsc{[print]}}{\effect{\userLan{s}}}}
%        \hypo{\assert{\effect{\userLan{s}}}{\textsc{(cbn-beta)}}{\dupliEffect{op}}}
        \hypo{
        \begin{array}{l}
%        \textit{Below, eff = \effect{} and dup = \dupliEffectName}\\
%        \textit{Below, noDupBeta = }\dupliEffect{\textsc{[beta]}}\\
        \navy{
       \assert{}{\textsc{[print]}}{\effect{\userLan{2}}}
       }
       \\
        \magenta{
       \assert{\effect{\userLan{2}}}{\textsc{[beta]}}{\textit{noDupBeta}}
       }
       \\
        \assert{\textit{noDupBeta}}{r_1}{\textit{noDupBeta}} ~
        \cdots ~
        \assert{\textit{noDupBeta}}{r_n}{\textit{noDupBeta}}
%        \assert{\textit{noDupBeta}}{r_1}{\textit{noDupBeta}}\\
%        \ldots\\
%        \assert{\textit{noDupBeta}}{r_n}{\textit{noDupBeta}}
        \end{array}
        }
%        \hypo{F}
        \infer1[\textsc{(inf)}]{\assert{}{\textsc{[print]} \app \textsc{[beta]} \app r_1 \cdots \cdots r_n}{\textit{noDupBeta}}}
%        \infer4{\assert{}{\textsc{[print]} \app \textsc{(cbn-beta)} \app r_1 \cdots \cdots r_n}{}}
        \infer1[\textsc{(perm-r)}]{\assert{}{r_1 \cdots \textsc{[print]} \cdots \textsc{[beta]}\cdots r_n}{\textit{noDupBeta}}}
%        \infer2{\assert{}{(op \app t_1, \ldots, t_n), s_1, \ldots, s_m \step t, s'_1, \ldots, s'_m}{\key{effect}{s}}
        \infer2[\textsc{(lang)}]{\assert{}{(G,I)}{\dupliEffect{\textsc{[beta]}}}}
    \end{prooftree}
    }
    \]

where \textit{noDupBeta = }\dupliEffect{\textsc{[beta]}}. 
The premises in color are derived as follows. 
Below, \textsc{(effectual-args)} satisfies the central premise of \textsc{(consequence)} 
and $\magenta{\effect{}}$ is forgotten in the way shown in Appendix \ref{forget}. 
%and a formula is forgotten in the way shown in Appendix \ref{forget}. 
\begin{gather*}
\navy{
\ninference{effect}
{ s_1 \not= (s_1 + \text{\quoting{$\returnSymbol$}} + s_2)
}
{\assert{}{\textsc{[print]}:\print{s_2}, s_1 \step   \unitE, s_1 + \text{\quoting{$\returnSymbol$}} + s_2}{\effect{2}}}
}
\\[1ex]
\magenta{
\ninference{effectual-args \& consequence}
{
%\effect{i} \in P \\
%v\in \ineffect 
%\\ 
%~\land~
%v \grammarDerivationG (\lambda x:T. e_1) ~\land~ v \grammarDerivationG v
v\in \ineffect 
~\land~
v \grammarDerivationG (\lambda x:T. e_1) \textit{ (* \quoting{implies} vacuously satisfied *)}
\\\\
v\in \ineffect 
~\land~
v \grammarDerivationG v \textit{ (* \quoting{implies} vacuously satisfied *)}
}
{\assert{\effect{\userLan{2}}}{\textsc{[beta]}:(\lambda x:T. e)\app v, s \step  e[v/x], s}{\dupliEffect{\textsc{[beta]}}}}
}
\end{gather*}

\paragraph{Issue 1b: \textsc{[beta]} is Not Ctx-compliant.}
%Having debugged our function application, we have  
With \textsc{[beta]}, we have 
$\expOne\centernot\step$
because the argument $(\print \text{\quoting{\textit{Performing a parameter passing}}} ; 2)$ does not evaluate. 
Indeed, the only evaluation context for application is $(E \app e)$. 
We see that \textsc{(ctx-compliant)} cannot be applied for \textsc{[beta]} with $\ctxPositions{\userLan{E}}{app}{\{1\}}$: 
%We see that \textsc{(ctx-compliant)} cannot be applied for \textsc{[beta]} with precondition $\ctxPositions{\userLan{E}}{app}{\{1\}}$. 
\[
\ninference{ctx-compliant}
{ v \grammarDerivationG (\lambda x:T. e) \myimplies 1 \in \{1\} \textit{ (* first arg of $(\lambda x:T. e)\app v$ *)}
\\\\
%v \Rightarrow v \Rightarrow \redd{2 \in \{1\}} 
v \grammarDerivationG v  \redd{~\land~ 2 \not\in \{1\}} 
\textit{ (* second arg of $(\lambda x:T. e)\app v$ *)}
}
{\assert{\ctxPositions{\userLan{E}}{app}{\{1\}}}{\textsc{[beta]}:(\lambda x:T. e)\app v, s \step  e[v/x], s}{ \app \ldots \app }}
\]

We can debug our language by adding the evaluation context $(v\app E)$ (and error context $(v\app F)$ for completeness). 
%We can then derive the following.
We can then derive the following, 
where $ctxApp = \ctxPositions{\userLan{E}}{app}{\{1,2\}}$ and $\textit{ctxOKBeta} = \reachable{\textsc{[beta]}}$.

\[
%\hspace{-0.9em}
{\small
\begin{prooftree}
%        \hypo{A}
%        \infer1{B}
%        \hypo{C}
        \hypo{
        \navy{
        \assert{}{G}{ \textit{ctxApp}}
        }
        %\end{array}
        }
%        	\rewrite{\color{lightblue}\box\treebox}
%        \hypo{\assert{}{\textsc{[print]}}{\effect{\userLan{s}}}}
%        \hypo{\assert{\effect{\userLan{s}}}{\textsc{(cbn-beta)}}{\dupliEffect{op}}}
        \hypo{
        \begin{array}{l}
                \magenta{
       \assert{\textit{ctxApp}}{\textsc{[beta]}}{ \textit{ctxOKBeta}}
       }\\
        \assert{ \textit{ctxOKBeta}}{r_1}{ \textit{ctxOKBeta}}
        ~
        \cdots ~
        \assert{ \textit{ctxOKBeta}}{r_n}{ \textit{ctxOKBeta}}
        \end{array}
        }
%        \hypo{F}
%        \infer1{\assert{}{\textsc{[print]} \app \textsc{[beta]} \app r_1 \cdots \cdots r_n}{\dupliEffect{app}}}
%        \infer4{\assert{}{\textsc{[print]} \app \textsc{(cbn-beta)} \app r_1 \cdots \cdots r_n}{}}
        \infer1{\assert{\textit{ctxApp}}{\textsc{[beta]} ~ r_1   \cdots  r_n}{{\textit{ctxOKBeta}}}}
                \infer1[\textsc{(perm-r)}]{\assert{\textit{ctxApp}}{r_1   \cdots \textsc{[beta]}\cdots r_n}{{\textit{ctxOKBeta}}}}
%        \infer2{\assert{}{(op \app t_1, \ldots, t_n), s_1, \ldots, s_m \step t, s'_1, \ldots, s'_m}{\key{effect}{s}}
        \infer2{\assert{}{(G,I)}{\reachable{\textsc{[beta]}}}}
    \end{prooftree}
}
\]
%\textit{where} $ctxApp = \ctxPositions{\userLan{E}}{app}{\{1,2\}}$, $\textit{ctxCompl}\beta = \reachable{\textsc{[beta]}}$.
%\textit{where} $ctxApp = \ctxPositions{\userLan{E}}{app}{\{1,2\}}$, $\textit{ctxOKBeta} = \reachable{\textsc{[beta]}}$.

The premises in color are derived as follows. 
\[
\navy{
\begin{prooftree}
%        \hypo{\assert{}{s_1}{}}
	\hypo{(E\app e).\getArgs{E} \cup (v\app E).\getArgs{E} = \{1\} \cup \{2\} = \{1,2\}}
	\rewrite{\color{lightblue}\box\treebox}
%        \hypo{\assert{}{\textsc{[print]}}{\effect{\userLan{s}}}}
%        \hypo{\assert{\effect{\userLan{s}}}{\textsc{(cbn-beta)}}{\dupliEffect{op}}}
        \infer1[\navy{\textsc{(inductive)}}]{
        \begin{array}{l}
        \qquad
        \navy{
        \assert{}{(\textsf{EvalCtx} \app E ::= \cdots \mid E\app e \mid v\app E \cdots )}{\textit{ctxApp}}
        %{\ctxPositions{\userLan{E}}{app}{\{1,2\}}}
        }
        \\[0.5ex]
%\qquad
\qquad       
\assert{\textit{ctxApp}}{g_1}{\textit{ctxApp}}~
        \cdots~
       \assert{\textit{ctxApp}}{g_n}{\textit{ctxApp}}
%       \assert{\ctxPositions{{E}}{app}{\{1,2\}}}{g_1}{\ctxPositions{{E}}{app}{\{1,2\}}}\\
%        \ldots\\
%       \assert{\ctxPositions{{E}}{app}{\{1,2\}}}{g_n}{\ctxPositions{{E}}{app}{\{1,2\}}}\\
        \end{array}
        }
        \infer1[\textsc{(grammar)}]{\assert{}{(\textsf{EvalCtx} \app {E} ::=  \cdots) \app g_1 \app \cdots \app g_n}{\textit{ctxApp}}}%{\ctxPositions{{E}}{app}{\{1,2\}}}}
%        \hypo{F}
%        \infer1{\assert{}{\textsc{[print]} \app \textsc{[beta]} \app r_1 \cdots \cdots r_n}{\dupliEffect{app}}}
%        \infer4{\assert{}{\textsc{[print]} \app \textsc{(cbn-beta)} \app r_1 \cdots \cdots r_n}{}}
%        \infer2{\assert{}{(op \app t_1, \ldots, t_n), s_1, \ldots, s_m \step t, s'_1, \ldots, s'_m}{\key{effect}{s}}
        \infer1[\textsc{(perm-g)}]{\assert{}{g_1   \cdots (\textsf{EvalCtx} \app {E} ::=  \cdots)\cdots g_n}{\ctxPositions{{E}}{app}{\{1,2\}}}}
    \end{prooftree}
}    
    \]
\[
\magenta{
\ninference{ctx-compliant \& consequence}
{ 
P = {\ctxPositions{\userLan{E}}{app}{\{1,2\}}} \\ {\ctxPositions{\userLan{E}}{app}{\{1,2\}}} \in P \\\\
v \grammarDerivationG (\lambda x:T. e) \myimplies 1 \in \{1,2\} \\
v \grammarDerivationG v \myimplies 2 \in \{1,2\} 
}
{\assert
%{\ctxPositions{\userLan{E}}{app}{\{1,2\}}}
{P}
{\textsc{[beta]}:(\lambda x:T. e)\app v, s \step  e[v/x], s}{\reachable{\textsc{[beta]}}}}
}
\]

%Thanks to our fixes, $\expOne, \text{\quoting{}} \step^{*} \app 1, \text{\quoting{\textit{Performing a parameter passing}}}$. 
After this fix, we have $\expOne \step^{*} \app 1, \text{\quoting{\textit{Performing a parameter passing}}}$. 

%\textit{On fixing proof derivations: } 
%Notice that the addition of $(v\app E)$ (and $(v \app F)$) does not make $\dupliEffect{\textsc{[beta]}}$ invalid. 
%The addition of $(v\app E)$ (and $(v \app F)$) does not make $\dupliEffect{\textsc{[beta]}}$ invalid. 
This fix does not make $\dupliEffect{\textsc{[beta]}}$, previously derived, invalid. 
However, its proof derivation needs to use the new (syntactically different) grammar. 
(The same is true for program logics: 
%Innocuous syntactic changes to the program mean that previous derivations are invalid unless they use the new program.
%Irrelevant syntactic changes mean that derivations become invalid unless they use the new program.) 
Innocuous changes to the program mean that previous proof derivations do not work as they are but must use the new program even if the structure of the proof does not change.) % as they are.) % unless they use the new syntax.) 

\paragraph{Issue 2: Error Handler May Be Ignored.} 
%We now test $\runningExample$ for division errors and for catching those errors with \key{try}. 
%Passing $0$ to our function with $\expTwo = ((\lambda x: \Float. \app x \division x)  \app 0)$ works, i.e., $\expTwo, \text{\quoting{}} \step^{*} \app \key{error},  \text{\quoting{}}$. 
%However, if we try to catch that error with 
%If we try to catch errors with \key{try}, we discover the following. \\
If we test \key{try}, we discover the following. \\

%
%\begin{gather*}
$
\begin{array}{ll}
\expTwo & = \key{try} \app 2\division 0 \app \key{with}\app (\print \text{\quoting{\textit{Division by $0$ occurred}}} ; \key{error}), \,  \text{\quoting{\,\!}}\\
%\expTwo = &\key{try} \app 2\division 0 \app \key{with}\app (\print \text{\quoting{\textit{Division by $0$ occurred}}} ; \key{error}), \,  \text{\quoting{\,\!}}\\
&\step^{*} 
%\\
%&
\key{try} \app \key{error} \app \key{with}\app (\print \text{\quoting{\textit{Division by $0$ occurred}}} ; \key{error}), \,  \text{\quoting{\,\!}}\\
&\step 
%\\
%&
\key{error}, \,  \text{\quoting{\,\!}}
\end{array}
$
%\end{gather*}

The last step disregarded the \key{with}-clause of \key{try}. 
Rule \textsc{[err-ctx]} detected the error with error context $\try{F}{e}$ and failed the computation. 
If we attempted to apply the proof rule \textsc{(error-handler)} to \key{try} after having derived the error contexts of \key{try}, i.e., $\ctxPositions{\userLan{F}}{try}{\{1\}}$, 
%and that the reduction rule \textsc{(err)} is reachable, 
and also \reachable{\textsc{[err]}}, 
we would fail because the red premise below holds whereas we had to satisfy the condition that such argument was \emph{not} subject to an error context.  
\[
\ninference{error-handler}
{ \reachable{\textsc{[err]}} \in P \\  \ctxPositions{\userLan{F}}{try}{\{1\}} \in P\\\\
err \grammarDerivationG \key{error} 
%\\\\
\\
\redd{1 \in  \{1\}}
}
{\assert{P}{\textsc{[err]}:\try{\key{error}}{e}, s  \step  e, s}{ \app \ldots \app }}
\]

%Unless , it is generally correct that evaluation contexts are also error
We can fix this by removing that error context. 
This leaves \key{try} with no error contexts declared in \textsf{ErrCtx}. 
%We can then derive the following. 
We can then derive the following, where 
$\textit{ctxTry} = \ctxPositions{\userLan{E}}{try}{\{1\}}$, $\textit{tryF} = \ctxPositions{\userLan{F}}{try}{\{\}}$, and 
$\errHandlerShort{try}{1} = \errHandler{try}{1}$.
\[
%\hspace{-1.2em}
{\small
\begin{prooftree}
%        \hypo{A}
%        \infer1{B}
%        \hypo{C}
%        \hypo{\assert{}{G}{\arrayTwo{
%        \ctxPositions{\userLan{E}}{try}{1}}
%        {\ctxPositions{\userLan{F}}{try}{1}}
%        }}
        \hypo{\assert{}{G}{\textit{ctxTry} \land \textit{tryF}\,}
        }
        	\rewrite{\color{lightblue}\box\treebox}
%        \hypo{\assert{}{{s_1   \cdots (\textsf{EvalCtx} \app \langVar{E} ::=  \cdots})}{\ctxPositions{\userLan{E}}{try}{1}}}
%        \hypo{\assert{}{\textsc{[print]}}{\effect{\userLan{s}}}}
%        \hypo{\assert{\effect{\userLan{s}}}{\textsc{(cbn-beta)}}{\dupliEffect{op}}}
        \hypo{
        \begin{array}{l}
        \magenta{
       \assert{\textit{ctxTry} \land \textit{tryF}\,
       }
       {\textsc{[err]}}
       {\errHandlerShort{try}{1}\,}
       }
%       \\[1.5ex]
       \\
        \assert{\errHandlerShort{try}{1}\,}{r_1}{\errHandlerShort{try}{1}\,} ~ 
        \cdots ~
        \assert{\errHandlerShort{try}{1}\,}{r_n}{\errHandlerShort{try}{1}\,}
%        \assert{\errHandlerShort{try}{1}\,}{r_1}{\errHandlerShort{try}{1}\,}\\
%        \ldots\\
%        \assert{\errHandlerShort{try}{1}\,}{r_n}{\errHandlerShort{try}{1}\,}
        \end{array}
        }
%        \hypo{F}
%        \infer1{\assert{}{\textsc{[print]} \app \textsc{[beta]} \app r_1 \cdots \cdots r_n}{\dupliEffect{app}}}
%        \infer4{\assert{}{\textsc{[print]} \app \textsc{(cbn-beta)} \app r_1 \cdots \cdots r_n}{}}
        \infer1[\!\textsc{(inf)}]{
        \assert{ \textit{ctxTry} \land \textit{tryF}\, }
        {\textsc{[err]} \app r_1 \cdots r_n}{\errHandlerShort{try}{1}\,}}
        \infer1[\textsc{(perm-r)}]{
        \assert{ \textit{ctxTry} \land \textit{tryF}\, }
        {r_1 \cdots \textsc{[err]}  \cdots r_n}{\errHandlerShort{try}{1}\,}}
        \infer2[\textsc{(lang)}]{\assert{}{(G,I)}{\errHandler{try}{1}\,}}
    \end{prooftree}
}
    \]
    
%\textit{where} $\textit{ctxTry} = \ctxPositions{\userLan{E}}{try}{\{1\}}$, $\textit{tryF} = \ctxPositions{\userLan{F}}{try}{\{\}}$, and 
%$\errHandlerShort{try}{1} = \errHandler{try}{1}$.
We need \textit{ctxTry} to derive $\reachableName$ for \textsc{[err]}, which is required by \textsc{(error-handler)}. 
This step is not explicit above because it is within an application of \textsc{(iterate)}, which we show below. 

The premises in color are derived as follows. 
The derivation of \textit{ctxTry} follows the same lines as that of \textit{ctxApp}, and we show it in Appendix \ref{ctxTry}. 
Then, the derivation of \textit{tryF} begins with \textit{ctxTry} as an already derived precondition. 
\[
\label{pageOfctxTry}
\navy{
\ninference{inductive}
	{
	\{\} \subseteq \{1 \ldots n\}  ~~ \textit{(* $I = \{\}$ as try is not in any $t_i$ *)}\\\\
%	\forall j\in I, t_j.\key{op} = op \\ \forall j\in (\{1,\ldots, n\} - I), t_j.\key{op} \not= op\\\\
	I' = \{\} ~~ \textit{(* as $\bigcup$ for $j\in \{\}$ is vacuously $\{\}$ *)} 
	%\\\\
%	\assert{P_0}{\userLan{cname \app \langVar{X} ::= t_1}}{P_1} \\\\
%	\assert{P_1}{\userLan{cname \app \langVar{X} ::= t_2}}{P_2} \\\\
%	\ldots \\\\
%	\assert{P_{n-1}}{\userLan{cname \app \langVar{X} ::= t_n}}{P_n} 
	} 
	{\assert{\textit{ctxTry}\,}{(\textsf{ErrCtx} \app {F} ::=  t_1\app  \mid \cdots \app \mid \app  t_n)}
	{  \textit{ctxTry} \land \app \ctxPositions{\userLan{F}}{try}{\{\}}
%                       \arrayTwo{
%        \textit{reach}
%        }
%        {
%        \land \app \ctxPositions{\userLan{F}}{try}{\{\}}
%        }
        }
	}%\,\,\textsc{(t-head)}
}
\]
\[
\magenta{
\ninference{iterate}
 {
 \assert
{\textit{ctxTry} \land \textit{tryF}\,}
%{\textit{reach} \land \textit{tryF}}
{\textsc{[err]}}
{
\begin{array}{l}
\textit{ctxTry} \\
\land~ \textit{tryF}\,  \\
\land ~ \textit{ctxOKTry}
\end{array}
}
\\
\assert
%{\textit{ctxTry} \land \textit{tryF} \land \textit{ctxOKTry}}
{
\begin{array}{l}
\textit{ctxTry} \\
\land~ \textit{tryF}\,  \\
\land ~ \textit{ctxOKTry}
\end{array}
}
{\textsc{[err]}}
{\errHandlerShort{try}{1}\,}
}
{\assert
{\textit{ctxTry} \land \textit{tryF}\,}
%{\textit{reach} \land \textit{tryF}}
{\textsc{[err]}}{\keydash{error}{handler}(try,1)}}
}
\]\label{iterate}

where $\textit{ctxOKTry} =\reachable{\textsc{[err]}}$. 
The derivation of \textit{ctxOKTry} (first premise of \textsc{(iterate)}) follows the same lines as that of \textit{ctxOKBeta} for application, 
and we show it in Appendix \ref{ctxOKTry}. 
%The second premise, which derives $\errHandler{try}{1}$, is derived as follows. 
The second premise is derived as follows. 
\[
\magenta{
\ninference{error-handler \& consequence}
{ 
P = \textit{ctxTry} ~\land ~\ctxPositions{\userLan{F}}{try}{\{\}} ~\land ~\reachable{\textsc{[err]}} \\\\
\reachable{\textsc{[err]}} \in P \qquad
\ctxPositions{\userLan{F}}{try}{\{\}} \in P \\
err \grammarDerivationG \key{error} 
%%\\\\
\\
1 \not\in  \{\}
}
{\assert
{P}
{\textsc{[err]}: \try{\key{error}}{e}, s  \step  e, s}{\errHandler{try}{1}\,}}
}
\]
After this fix, we have $\expTwo \step^{*} \key{error}, \text{\quoting{\textit{Division by 0 occurred}}}$.

\paragraph{Issue 3: Typing Rule Does Not Respect Contravariance.}

Program $\expOne$ does not type check successfully. 
(To recall, $\expOne = ((\lambda x: \Float. \app x \division x)  \app\app (\print \cdots ; 2)), \,  \text{\quoting{\textit{\,\!}}}$, where we omitted the string for brevity.) %[2ex]
%(To recall, \expOne = ((\lambda x: \Float. \app x \division x)  \app\app (\print \text{\quoting{\textit{Performing a parameter passing}}} ; 2))\\%[2ex]
The reason is that \textsc{(t-app-bad)} requires the type of the domain of the function to be a subtype of the type of argument, 
which is not consistent with the contravariance of function types. 
 %i.e., with the premise $T_1 <: T_3$. 
This means that we need $\Float <: \Int$ in our example, which does not hold. 
%and it is ordinarily inconsistent with  contravariance in function type. 
If we try to apply \textsc{(contra-respecting)} to \textsc{(t-app-bad)} after having derived that the domain of function types is contravariant ($\contravariant{\to}{\{1\}}$), we would fail:  %because of the fact formulae in red below, which contradict \textsc{(contra-respecting)}'s check that contravariant types cannot be used at the lefthand side of $<:$. 
\[
\ninference{contra-respecting}
{
P =  \contravariant{\to}{\{1\}} \\ \contravariant{\to}{\{1\}} \in P\\\\
%T_3 <: T_1 \Rightarrow \Gamma \typeOf \app e_1 : T_1\to T_2 \Rightarrow (T_1\to T_2).args(T) = \{1\} \subseteq \{1\} 
f = \Gamma \typeOf \app e_1 : T_1\to T_2 
%\Rightarrow 
%\redd{\lnot\exists \textit{ premise }T_1 <: T_3}}
\redd{~\land~1\in\{1\}~\land~\exists \textit{premise }T_1 <: T_3}}
{\assert{P}
{\textsc{[t-app-bad]}: \inference	{
	 \Gamma \typeOf \app e_1 : T_1\to T_2 &
	 \Gamma \typeOf \app e_2 : T_3 & T_1 <: T_3}  	
	{  \Gamma \typeOf \app e_1\app e_2 : T_2}
}
{ \app \ldots \app }}
\]
\noindent We can fix our language by replacing \textsc{[t-app-bad]} with the correct typing rule: 
%\noindent We can fix our language by replacing \textsc{[t-app-bad]} with the correct typing rule \textsc{[t-app]}. 
%Then, we can apply derive $\variancePres{\textsc{[t-app]}}{\to}$. 
\[
%\ninference{t-app}
\textsc{[t-app]:}\inference
%\inference
	{
	 \Gamma \typeOf \app e_1 : T_1\to T_2 &
	 \Gamma \typeOf \app e_2 : T_3 & \HI{$T_3 <: T_1$}  	
	} 
	{  \Gamma \typeOf \app e_1\app e_2 : T_2}
\]
%We can then apply \textsc{(contra-respecting)} and derive the following. 
After this, $\expOne$ type checks successfully. 
%We also can derive the following. \\
We also can derive the following, where 
\textit{contra =}$\contravariant{\to}{\{1\}}$ and $\textit{contraOK} = \variancePres{\textsc{[t-app]}}{\to}$. 
%\noindent \!\textit{where contra =}$\contravariant{\to}{\{1\}}$, $\textit{contraOK} = \variancePres{\textsc{[t-app]}}{\to}$. 
%\\
%(The premises in color are derived immediately afterwards.)
\[
%\hspace{-0.8em}
\begin{prooftree}
%        \hypo{A}
%        \infer1{B}
%        \hypo{C}
        \hypo{\assert{}{G}{}}
%        \hypo{\assert{}{\textsc{[print]}}{\effect{\userLan{s}}}}
%        \hypo{\assert{\effect{\userLan{s}}}{\textsc{(cbn-beta)}}{\dupliEffect{op}}}
        \hypo{
        \begin{array}{l}
        \navy{
        \assert{}{\textsc{[s-arrow]}}{\textit{contra}}
        }
        \\
        \magenta{
        \assert{\textit{contra}}{\textsc{[t-app]}}{\textit{contraOK}}
        }
        \\
        \assert{\textit{contraOK}}{r_1}{\textit{contraOK}} ~
        \cdots ~
        \assert{\textit{contraOK}}{r_n}{\textit{contraOK}}\\
%        \assert{\textit{contraOK}}{r_1}{\textit{contraOK}}\\
%        \ldots\\
%        \assert{\textit{contraOK}}{r_n}{\textit{contraOK}}\\
        \end{array}
        }
%        \hypo{F}
%        \infer1{\assert{}{\textsc{[print]} \app \textsc{[beta]} \app r_1 \cdots \cdots r_n}{\dupliEffect{app}}}
%        \infer4{\assert{}{\textsc{[print]} \app \textsc{(cbn-beta)} \app r_1 \cdots \cdots r_n}{}}
        \infer1[\textsc{(inf)}]{\assert{}{\textsc{[s-arrow]} \app \textsc{[t-app]} \app r_1   \cdots r_n}{\textit{contraOK}}}
        \infer1[\textsc{(perm-r)}]{\assert{}{r_1   \cdots \textsc{[t-app]}\cdots \textsc{[sub-arrow]}\cdots r_n}{\textit{contraOK}}}
%        \infer2{\assert{}{(op \app t_1, \ldots, t_n), s_1, \ldots, s_m \step t, s'_1, \ldots, s'_m}{\key{effect}{s}}
        \infer2[\textsc{(lang)}]{\assert{}{(G,I)}{\variancePres{\textsc{[t-app]}}{\to}}}
    \end{prooftree}
    \]
    
%\noindent \!\textit{where contra =}$\contravariant{\to}{\{1\}}$, $\textit{contraOK} = \variancePres{\textsc{[t-app]}}{\to}$. \\
The premises in color are derived as follows. 
Below, $\{1\}$ is the result of $\bigcup_{(T_i' <: T_i) \in \widetilde{f}} ~\{i\}$ where only $T_1' <: T_1$ contributes. 
%
%The premises in color are derived as follows. 
%($\bigcup_{(T_i' <: T_i) \in \widetilde{f}} ~\{i\}=\{1\}$ where only $T_1' <: T_1$ contributes.)
\[
\navy{
\ninference{contravariant}
%{\{1\} ~~\textit{(* $h(T_1' <: T_1) \cup h(T_2 <: T_2') = \{1\} \cup \emptyset$ *)}}
{}
{\assert{}{ \textsc{[s-arrow]}:\inference{T_1' <: T_1 & T_2 <: T_2'}{T_1 \to T_2 <: T_1' \to T_2'}}{ \contravariant{\to}{\{1\}}}}
}
%\ninference{contravariant}  \contravariant{c}{n}\mid \variancePres{rn}{c}
%{}
%{\assert{P}{\inference{\ldots \app T_i' <: T_i \app \ldots}{(c \app T_1 \app \ldots \app T_n) <: (c \app T_1' \app \ldots \app T_n') }}{P \land \key{contravariant}(c,i)}}
%
\]
%
%where $\{1\}$ is the result of $\bigcup_{(T_i' <: T_i) \in \widetilde{f}} ~\{i\}$ where only $T_1' <: T_1$ contributes. 
%
\[
\magenta{
\ninference{contra-respecting \& consequence}
{
P =  \contravariant{\to}{\{1\}} \\ \contravariant{\to}{\{1\}} \in P\\\\
%T_3 <: T_1 \Rightarrow \Gamma \typeOf \app e_1 : T_1\to T_2 \Rightarrow (T_1\to T_2).args(T) = \{1\} \subseteq \{1\} 
f = \Gamma \typeOf \app e_1 : T_1\to T_2 \myimplies \lnot\exists \textit{premise }T_1 <: t_i'}
{\assert{P}
{\textsc{[t-app]}: \inference	{
	 \Gamma \typeOf \app e_1 : T_1\to T_2 &
	 \Gamma \typeOf \app e_2 : T_3 & T_3 <: T_1}  	
	{  \Gamma \typeOf \app e_1\app e_2 : T_2}
}
{\textit{contraOK}}}
%{\variancePres{\textsc{[t-app]}}{\to}}}
}
\]
%where $\{1\}$ is the result of $\bigcup_{(T_i' <: T_i) \in \widetilde{f}} ~\{i\}$ where only $T_1' <: T_1$ contributes. 

%After this fix, $\expOne$ type checks successfully.  

%\noindent Also, having replaced \textsc{(t-bad-app)} with \textsc{(t-app)}, all our examples type check. 

%\paragraph{Limitations of Language Logics.} 

\paragraph{Implementation.}

We have implemented an automated prover for the logic $\langLogic$ in OCaml. 
The tool is called $\lna$ \cite{lna} and takes in input the elements of a statement: a precondition assertion, a language definition, and a postcondition. 
%Language definitions are a textual representation of operational semantics, an example is in Appendix \ref{lan}. 
Language definitions are given in a textual representation, an example is in Appendix \ref{toolLanguage}. 
%$\lna$ attempts to derive the statement based on the proof rules of $\langLogic$. 
%The output is either a proof derivation or an error message. 
%The output is a proof derivation or a \quoting{no proof found} message. 
The output is a proof derivation or a no-proof-found error message. 
%, if the tool could not prove the  statement. 

$\lna$ is a \emph{forward reasoner}. It starts from the input precondition (which ordinarily is $\{\}$) and analyzes all the grammar and inference rules. % in the order they appear in the language. 
%For each, $\lna$ tries all the proof rules that are applicable, once, to derive their assertions. 
For each, $\lna$ tries all the (base case) proof rules of Fig. \ref{fig:rulesRest}, once, to derive their assertions. 
These are a finite number of attempts. 
%These are finite attempts. 
After having analyzed the language, $\lna$ has accumulated all the assertions derived. 
%For each, $\lna$ tries to apply all the proof rules to derive their corresponding assertion, if applicable. These are finite applications. 
%Based on the rules of $\langLogic$, these are finite. 
%To avoid non-determinism in that process, $\lna$ does apply \textsc{perm-g} nor \textsc{perm-r}. 
To avoid non-determinism, we use \textsc{(consequence)} only at the end to single out the goal assertion, 
and we do not apply permutation rules 
%To avoid non-determinism, $\lna$ does not apply permutation rules 
except at the beginning to place subtyping rules before typing rules because \textsc{(contra-resp)} needs $\contravariantName$. 
Since \textsc{(error-handler)} needs $\reachableName$ \emph{of the same rule}, we always apply \textsc{(ctx-compliant)} before \textsc{(error-handler)}. % for each rule. 
%Also, \textsc{(error-handler)} needs $\reachableName$ \emph{of the same rule}, we therefore always apply \textsc{(ctx-compliant)} before \textsc{(error-handler)}. % for each rule. 
%Other proof rules may depend on \key{inductive} but this dependency is trivially fulfilled because the logic analyzes the grammar before inference rules. 
%Indeed, the only dependency in that regard is that subtyping rules must be analyzed before typing rules 
%$\lna$ reorders the inference rules in that way as first thing. %with one application of \textsc{}
%applies \textsc{perm-g} nor \textsc{perm-r}.There are two dependencies 
%, and $\langLogic$, specifically does so regardless if places them after typing rules. 
%\texttt{\{\} lang.lan \{\}}

%We have applied $\lna$ to the debugging of $\runningExample$ of Section \ref{examples}. 
We have applied $\lna$ to the debugging of $\runningExample$. 
We confirm that we can repeat the debugging journey of Section \ref{examples}. 
%(\textit{Issue 1a}): $\lna$ fails to derive the assertion $\dupliEffect{\textsc{[cbn-beta]}}$, so we fix the issue and $\dupliEffect{\textsc{[beta]}}$ succeeds. 
%(\textit{Issue 1a}): $\lna$ fails to derive $\dupliEffect{\textsc{[cbn-beta]}}$, but with \textsc{[beta]} we have that $\dupliEffect{\textsc{[beta]}}$ succeeds. 
(\textit{Issue 1a}): $\lna$ fails to derive $\dupliEffect{\textsc{[cbn-beta]}}$ but $\dupliEffect{\textsc{[beta]}}$ succeeds after we use \textsc{[beta]}. 
%(\textit{Issue 1a}): $\lna$ fails to derive $\dupliEffect{\textsc{[cbn-beta]}}$, so we fix the issue with \textsc{[beta]} and it succeeds.   
% after replacing \textsc{[cbn-beta]} with \textsc{[beta]}. 
(\textit{Issue 1b}): Then, $\reachable{\textsc{[beta]}}$ fails, but it succeeds after we fix the issue. 
%(\textit{Issue 1b}): Then, $\reachable{\textsc{[beta]}}$ fails, so we fix the issue and it succeeds.  
(\textit{Issue 2}): Then, $\errHandler{try}{1}$ fails, so we fix the issue and it succeeds.  
(\textit{Issue 3}): Finally, $\variancePres{\textsc{[t-app-bad]}}{\to}$ fails, but with \textsc{[t-app]} we have that $\variancePres{\textsc{[t-app]}}{\to}$ succeeds. % after our fix. 
%(\textit{Issue 3}): Finally, $\variancePres{\textsc{[t-app-bad]}}{\to}$ fails, so we fix the issue and $\variancePres{\textsc{[t-app]}}{\to}$ succeeds. % after our fix. 
The tools' proofs differ from those in this paper, as they accumulate all assertions, for example. % derived along the way. 
%The tools' proof derivations differ from those in this paper, as they accumulate all assertions. % derived along the way. 
%The tools' proof derivations differ, for example they carry all the assertions derived along the way. 

Our tool website carefully documents all these tests and contains others \cite{lna}. 
%Our tool website carefully documents all these experiments and contains others \cite{lna}. 

%(\textit{Issue 1a}): $\lna$ fails to derive $\dupliEffect{\textsc{[cbn-beta]}}$ 
%and succeeds with $\dupliEffect{\textsc{[beta]}}$ after our fix. % after replacing \textsc{[cbn-beta]} with \textsc{[beta]}. 
%(\textit{Issue 1b}): Then, $\reachable{\textsc{[beta]}}$ fails and succeeds after our fix. 
%(\textit{Issue 2}): Then, $\errHandler{try}{1}$ fails and succeeds after our fix. 
%(\textit{Issue 3}): Finally, $\variancePres{\textsc{[t-app-bad]}}{\to}$ fails and $\variancePres{\textsc{[t-app]}}{\to}$ succeeds after our fix. 
%Our website carefully documents these experiments \cite{lna}. 

%$\lna$ fails to derive $\dupliEffect{\textsc{[cbn-beta]}}$ 
%We confirm that $\lna$ fails to derive $\dupliEffect{\textsc{[cbn-beta]}}$ for $\runningExample$, and succeeds with $\dupliEffect{\textsc{[beta]}}$  
%Then, the tool failed $\reachable{\textsc{[beta]}}$ but succeeded after our fix. 

\section{Comparison with Program Logics}\label{comparison}\label{comparison}

%There are several differences between program logics and language logics as presented here. 
Program logics define the proof rules of $\assert{P}{c}{Q}$ by induction on $c$. % (besides \quoting{structural rules}).  
That is, there is a proof rule for \key{if}, assignment, \key{while}, and so on. 
These proof rules are based on the invariants that we learn after executing a command. 
In language logics: 
%What do we learn from adding the inference rule $\inference{f_1 \app \cdots \app f_n}{f}$ into a language? 
What do we learn from adding the inference rule $f_1 \land f_2 \app \cdots \app f_n \implies f$ 
(here in implicational notation)
into a language? 
Akin to adding a formula into a theory, we learn the formula itself. 
A same approach does not seem to be helpful in language logics. 
Therefore, our proof rules are based on the structure of $\LangDef$ only to traverse the language and reach grammar rules $g$ and inference rules $r$. 
For $g$ and $r$, our proof rules are defined by cases on the grammar of assertions. 
%A same view does not seem to be helpful in language logics, 
%therefore our proof rules are defined by cases on the assertions. 
%thus our proof rules are defined by cases on the assertions $P$. 
%therefore our proof rules are defined by cases on the assertions $P$. 

There are three major implications from designing the proof rules around assertions. 
1) When program logics analyze a command $(op \app \widetilde{t})$, an assertion $\assert{P}{(op \app \widetilde{t})}{Q}$ always exists. %, i.e., they always produce an invariant. 
%This is not true in language logics, 
This would not be true in language logics, 
therefore we added \textsc{($\mathcal{X}$-neutral)}. 
%2) Proof rules in program logics tend to 
2) Program logics tend to derive a postcondition that exhaustively characterizes the invariants learned from executing a command. On the contrary, language logics focus on one aspect at a time and therefore we added \textsc{(iterate)}. % to keep the focus of the analysis over the same subject. 
3) Lacking that exhaustive character for postconditions, language logics face challenges w.r.t. soundness and completeness (See Sect. \ref{limitations}). 
%\noindent (Also notice that we have not used negated assertions $\lnot P$ in our examples. %we have not used it. 
(Another remark is that we have not used negated assertions $\lnot P$ in our examples. %we have not used it. 
Appendix \ref{negation} provides an example of a possible proof rule that involves negation.) 

%in this respect, as we discuss in the next section. 
%3) The exhaustive character of program logics postconditions enables soundness and completeness results. 
%Language logics meet challenges in this respect, as we discuss in the next section. 

%n of $\textsf{EvalCtx} \app E ::= \ldots \mid   (v\app E)$ breaks, strictly speaking,  but also in program logic if you change the program then the derivation proofs break. 

%\section{Limitations of $\langLogic$ and of Language Logics}\label{limitations}
\section{Limitations of Language Logics}\label{limitations}

Proof rules of $\langLogic$ derive assertions based on syntactic patterns. 
We therefore have difficulties with soundness and completeness, which we do not offer. 
%Proof rules of $\langLogic$ derive assertions based on detecting common syntactic patterns. 
%We therefore have difficulties with soundness and completeness. 
%We therefore have a problem with both soundness and completeness. 

%\paragraph{Soundness Challenge} 
\fakePar{Soundness Challenge} 
\textsc{(effectual-args)} only considers explicit duplication or substitution. 
However, if a reduction rule makes use of a term linearly but in the context of a user-defined operation that acts like substitution, 
%However, if a reduction rule makes use of an argument linearly though with a user-defined operation that acts like substitution, 
%However, if a reduction rule makes use of an argument linearly though in the context of a user-defined operation that acts like substitution, 
the rule unsoundly derives a $\dupliEffectName$ assertion. 
%Deriving that \textsc{(beta)} is reachable in $\langLogic$ may be of no use if progress and preservation do not hold, 
%$(e_1 \app e_2)$ may get stuck regardless. % without reaching the form $((\lambda x.e) \app v)$. % regardless of the existence of evaluation contexts. 
%Similarly, proving the existence of evaluation contexts is of no use if rule \textsc{(ctx)} is broken.  % (say, it loops forever with $E[e] \step E[e]$ for example.) 
%work only provided that everything else is in place. 
%For example, the idea behind reachable may not $(e_1 \app e_2) \step^{*} ((\lambda x.e) \app v)$ may not 
%1) problem with contra when you have chain deceivePred T1 = T3, T3 <: T2. DETECT but it wasnt. 
% 
Also, we derive $\effect{}$ based on syntactic (in)equality between two states. 
If two distinct constants \key{a} and \key{b} for states are handled indistinguishably by a language, 
a reduction rule $t, \key{a} \step t', \key{b}$ unsoundly derives $\effect{}$ for the language. 
Similar remarks can be made for the other assertions of $\langLogic$. 
It is a challenge to design proof rules that guarantee a property while taking into account everything else that is in the language. 
Another challenge is in formulating a soundness statement for some aspects. 
%For example, it is not immediately clear what is the mathematical formula that establishes that a typing rule respects contravariance. 

%\paragraph{Completeness Challenge} 
\fakePar{Completeness Challenge} 
%The checks they perform are by no means complete. 
The checks that our proof rules perform are by no means complete. 
To make an example, 
consider a typing rule with three premises $\Gamma \typeOf e : T_1 \to T_2$, $(\textit{userDefinedEqual} \app T_{1} \app T_3)$, and $T_1 <: T_3$. 
%This rule simply type checks a function, as of function types, and 
The latter two premises are unnecessary given that \textit{userDefinedEqual} is equality (we assume), and subtyping is reflexive. \textsc{(contravariant-respecting)} detects $T_1$ at the left of $<:$ and does not derive $\variancePresName$. 
This is an example where a property holds but our proof system does not derive that. 
%To address this and similar cases, 
To address this, 
%\textsc{(contravariant-respecting)} 
the proof rule would have to thoroughly understand the relation between $T_1$ and $T_3$ (but notice that \textit{userDefinedEqual} may be complex at will, while encoding equality at the end.) 
Other assertions of $\langLogic$ suffer from similar issues. 
It is a challenge for language logics to completely characterize a property within proof rules. % syntactic means. 
%it is unclear it may not be possiblehow  to \emph{completely} characterize syntactically an aspect that we seek to derive. 

%1) problem with contra when arg is nested. OK but you did not catch. 
%2) problem with contra when you have chain T2 <: T1  T1 <: T3. OK but you did not catch. 
%problem to catch all... 

%\paragraph{All-encompassing Properties vs Small Questions} 
%\fakePar{All-encompassing Properties vs Small Questions} 
%\fakePar{All-encompassing Properties vs Selected Aspects} 
\fakePar{Challenge w.r.t. All-encompassing Properties} 
Following the distinction made at the beginning of Section 1, $\langLogic$ applies to \selected{} aspects of languages. %the language at hand. 
%(Program logics, too, have been used to selected invariants of small parts of programs.) 
%It is unclear how to develop language logics that can reason about 
It is unclear how to 
%develop language logics for
capture 
all-encompassing properties such as type soundness, strong normalization, and non-interference. 
Surely, $\langLogic$ must be extended. %We notice that inductive reasoning must be there, at least. 
For the progress theorem, for example, 
%from a well-typed $(e_1 \app e_2)$
%the proof derives the \quoting{progress} of $e_1$ and $e_2$ by induction (as they are well-typed) 
%before finding a reduction for all cases. 
a well-typed $(e_1 \app e_2)$ derives the \quoting{progress} of $e_1$ and $e_2$ by induction (as they are well-typed) 
before finding a reduction for all cases. 
Therefore, $\langLogic$ needs inductive reasoning, at least, which now is missing.  
%$\langLogic$ needs inductive reasoning, at least, which now is missing.  
%(Another remark is that we have not used negated assertions $\lnot P$ in our examples. %we have not used it. 
%Appendix \ref{negation} provides an example of a possible proof rule that involves negation.) 

%For example, checking that a reduction is type preservation may entail the assertions about 
%For example, the progress of $(e_1 \app e_2)$ depends on whether both $e_1$ and $e_2$ are well-typed, which then must be a precondition. 
%Furthering the comparison with program logics of Section \ref{comparison}, 
%we observe that program logics have been used for both full verification of programs (akin to global properties) and, say, selected loop invariants within large programs (akin to local aspects). 
%

%\redd{$\lna$ is publicly available and contains all our tests \cite{lna}.}

%\redd{$\lna$ is publicly available and contains all our tests \cite{lna}.}

\section{Related Work}\label{related}

%As we have discussed program logics, we address here related work in \emph{language-parametrized} language validation, i.e., where the language is given as input. 
We have discussed program logics in \S \ref{comparison}. 
%We have discussed program logics in Section \ref{comparison}. 
We address here related work in \emph{language-parametrized} language validation, i.e., where the language is given as input. 

%In this respect, a popular approach is that of testing random programs that are automatically generated 
%A popular approach is that of testing random programs that are automatically generated 
Much work is on testing random programs that are automatically generated 
from language definitions \cite{randomJudgements,mpsTesting,IsabelleTesting,PLTtesting}. %Babelsberg % \cite{randomJudgements,mpsTesting,IsabelleTesting}. % 
%This is the approach of many semantics engineering tools such as the $\mathbb{K}$ framework \cite{Rosu2010}, PLT Redex \cite{redex}, and MPS \cite{mps}, among others. 
%Similarly, 
Also, Robertson et al. \cite{Roberson} propose a model checking approach to testing the type soundness of a language in input. %at hand. 
These works differ from the approach with language logics in that they do not provide a proof derivation in a formal proof system.

%\textit{automated type soundness and testing BUT NO PROOF}
%\textit{automated type soundness WITH FORMAL PROOF but no flexibility in the property}
%\textit{flexibility in the property with lang-SQL but NO formal PROOF} no basis for certification. 

Some other works do establish a result by means of a proof. 
For example, \textsf{Veritas} \cite{Grewe:2015} and Twelf \cite{Pfenning1999} make use of automated theorem proving to establish the type soundness of a certain class of languages. %,Vampire2014and2015,Vampire2016    Twelf \cite{Pfenning1999,Schurmann00phd}
%Intrinsic typing \cite{ChurchTypes} leverages the meta-theoretic properties of a host type theory: If the evaluator type checks in such a type theory then the language is type soundness  \cite{Harper00,Altenkirch,Poulsen,Poulsen2,APPEL200795,thiemannSessionIntrinsic,BentonHKM12,Augustsson,intrinsicAlaCarte}.  
% shorter: \cite{Harper00,Poulsen,APPEL200795,thiemannSessionIntrinsic,BentonHKM12,intrinsicAlaCarte}.
Intrinsic typing leverages the meta-theoretic properties of a type theory to derive type soundness from a well-typed evaluator %ChurchTypes
\cite{Harper00,ChurchTypes,BentonHKM12}.  %Poulsen2, Altenkirch, Augustsson
%\cite{Harper00,Altenkirch,Poulsen,Poulsen2,APPEL200795,thiemannSessionIntrinsic,BentonHKM12,Augustsson,intrinsicAlaCarte}.  
%%Cimini et al. \cite{lnc,lncGit,tfp2019} propose an extrinsic type system that classifies parts of a language definition and imposes a language organization that ensures type safety. 
%%Cimini et al. \cite{lnc} generate a proof from an extrinsic type system analysis of a restricted class of language definitions. 
%$\lnp$ \cite{lnp} and \cite{lnc} apply the structure of a certain type soundness proof to a limited class of languages. 
The work in \cite{lnp} and \cite{lnc} applies/imposes the structure of a certain type soundness proof to a limited class of languages. %\cite{lnp} \cite{GalassoC24}
These works differ in that they specifically target type soundness. 
On the contrary, language logics can flexibly analyze various aspects. % of languages.  
%On the contrary, language logics can flexibly analyze various aspects of the language at hand. 
%derive all sorts of attributes of type constructors, operations, reduction rules, and so on. 
%(On the other hand, language logics face challenges with addressing type soundness, as discussed in Section \ref{limitations}.) 
%(On the other hand, language logics face challenges with type soundness, as discussed in Section \ref{limitations}.) 
%(On the other hand, it is unclear whether language logics can capture type soundness, as discussed in Section \ref{limitations}.) 
(On the other hand, it is unclear how to capture type soundness with language logics.) %, as discussed in Section \ref{limitations}.) 

%$\lnsql$ \cite{lnsqlTAP} offers some flexibility, as it interrogates languages with customizable queries. %\cite{lnsql,lnsqlTAP}
$\lnsql$ \cite{lnsql} can flexibly interrogate languages over various aspects with customizable queries. 
However, these queries do not provide a proof derivation.

\section{Conclusion}\label{conclusion}

In this paper, we have proposed the idea of \emph{language logics} as a counterpart of program logics in the context of language verification. 
%To demonstrate our approach, we have developed $\langLogic$, a language logic that can be used to analyze various aspects of programming languages. % of languages. 
We have developed an example of language logic, $\langLogic$, and we have applied it to the analysis of a faulty programming language and its debugging fixes. 
We have also implemented an automated prover for $\langLogic$ called $\lna$. % and reported on experiments.  % as an assertion prover called $\lna$. 
%We confirm that $\lna$ replicates the successes and failures of our debugging journey. % of the faulty language. 
We confirm that $\lna$ replicates the analyses that we performed on the faulty language. % in our debugging journey. % of the faulty language. 
Finally, we have provided a discussion of the limitations of language logics. % in the future. 

We acknowledge that language logics do not achieve language verification yet, 
%and we have discussed the main challenges in this regard. % our proof derivations do not formally establish a property. 
and we have offered in Section \ref{limitations} a discussion of the main challenges in this regard. 
%and Section \ref{limitations} offers a discussion of the main challenges in this regard. 
%and that our proof derivations do not formally establish a property. 
Nonetheless, we believe that this paper provides a strong first step towards adopting the methods of program logics for the analysis of languages. % language analysis. 

In the future, we would like to address the challenges described in Section \ref{limitations}, 
and hopefully elevate language logics to a full formal method for language verification. 
%especially insofar the soundness and completeness challenges is concerned. 
%In the future, we would like to address the limitations described in Section \ref{limitations}, 
%especially insofar the soundness and completeness challenges is concerned. 
%We would like to extend $\langLogic$ with other assertions in order to reason about other aspects of programming languages. 
We would like to extend $\langLogic$ with assertions on other aspects of programming languages. 
%We offer an example of extending $\langLogic$ in Appendix \ref{errorCtx}. 
%For example, it would be  related to non-interference, strong updates in the presence of aliasing, and 
%strong normalization, for example. 
Inspired by Software Foundations \cite{SFVolume2}, we would like to formalize $\langLogic$ in Coq so that proof derivations can be certified. 
%Certification in Coq, inspired by SF.
%Extend the logic to reason about aspects related to non-interference, strong normalization, 
%as well other unsound scenarios. 
%For example, 1) to detect that occurrence mutable dataypes in the presence of full polymorphism do not occur. 
%2) that bounds phenomenon with type projections of Amina and Tate's counterexample does not occur. 
%This may mean extending $\langLogic$ with new assertions. This is not dissimilar to program logics \key{while} had to be introduced. 
%Also, incorrectness logic.

$\lna$ is publicly available and contains all our tests \cite{lna}.

%% Bibliography
%\bibliographystyle{splncs04}
\bibliographystyle{plain}
\bibliography{all}

\newpage

\appendix

\section{Forgetting Assertions with \textsc{(consequence)}}\label{forget}
When $\assert{P}{\mathcal{X}}{P \land Q}$ can be derived, we can forget $P$ with
\[
\ninference{consequence}
	{
	P \Rightarrow P \\
	\assert{P}{\mathcal{X}}{P \land Q} \\
	P \land Q \Rightarrow Q \\
	} 
	{\assert{P}{\mathcal{X}}{Q}}% \,\,\textsc{(t-app)}
\]

%\section{Derivation of $\ctxPositions{E}{try}{\{1\}}$}\label{ctxTry}
%\section{Derivation of $\ctxPositions{E}{try}{\{1\}}$ and $\assert{}{G}{\textit{ctxTry} \land \textit{tryF}}$}\label{ctxTry}
\section{Derivation of $\assert{}{G}{\textit{ctxTry} \land \textit{tryF}\,}$ and \textit{ctxTry}}\label{ctxTry}

Recall $\textit{ctxTry} = \ctxPositions{{E}}{try}{\{1\}}$ and $\textit{tryF} = \ctxPositions{{F}}{try}{\{\}}$. 
The analysis of the whole grammar $G$ derives $\assert{}{G}{\textit{ctxTry} \land \textit{tryF}\,}$ as follows. 
\[
\begin{prooftree}
        \hypo{
        \begin{array}{c}
         \magenta{
         \assert{}{(\textsf{EvalCtx} \app E ::= \cdots)}{\textit{ctxTry}}
         }
         \\
         \navy{
         \assert{\textit{ctxTry}}{(\textsf{ErrCtx} \app F ::= \cdots)}{\textit{ctxTry} \land \textit{tryF}\,}
         }
         \\
       \assert{\textit{ctxTry} \land \textit{tryF}\,}{g_1}{\textit{ctxTry} \land \textit{tryF}\,}\\
        \ldots\\
       \assert{\textit{ctxTry} \land \textit{tryF}\,}{g_n}{\textit{ctxTry} \land \textit{tryF}\,}\\
        \end{array}
        }
        \infer1[\textsc{(gram.)}]{\assert{}{(\textsf{EvalCtx} \app {E} ::=  \cdots) \app (\textsf{ErrCtx} \app {F} ::=  \cdots) \app g_1 \app \cdots \app g_n}{\textit{ctxTry} \land \textit{tryF}\,}}
%        \hypo{F}
%        \infer1{\assert{}{\textsc{[print]} \app \textsc{[beta]} \app r_1 \cdots \cdots r_n}{\dupliEffect{app}}}
%        \infer4{\assert{}{\textsc{[print]} \app \textsc{(cbn-beta)} \app r_1 \cdots \cdots r_n}{}}
%        \infer2{\assert{}{(op \app t_1, \ldots, t_n), s_1, \ldots, s_m \step t, s'_1, \ldots, s'_m}{\key{effect}{s}}
%        \infer1[\textsc{(perm-g)}]{\assert{}{(\textsf{EvalCtx} \app {E} ::=  \cdots)  \app g_1 \app \cdots \app g_n \app (\textsf{ErrCtx} \app {F} ::=  \cdots)}{\textit{ctxTry} \land \textit{tryF}\,}}
        \infer1[\textsc{(perm-g)}]{\assert{}{g_1 \app \cdots \app g_n \app(\textsf{EvalCtx} \app {E} ::=  \cdots) \app (\textsf{ErrCtx} \app {F} ::=  \cdots)}{\textit{ctxTry} \land \textit{tryF}\,}}
            \end{prooftree}
    \]

where the statement above in blue color has been derived in Section \ref{examples}, page \pageref{pageOfctxTry}, and the statement in magenta color is derived as follows. 
\[
\magenta{
\ninference{inductive}
{(\try{{E}}{e}).\getArgs{E}  = \{1\} }
{\assert{}{(\textsf{EvalCtx} \app E ::= \cdots \mid \try{{E}}{e} \app \cdots )}{\ctxPositions{\userLan{E}}{try}{\{1\}}}}
}
\]

\section{Derivation of \textit{ctxOKTry}}\label{ctxOKTry}

Recall $\textit{ctxOKTry} = \reachable{\textsc{[err]}}$, $\textit{ctxTry} = \ctxPositions{{E}}{try}{\{1\}}$. 
%Recall that $\textit{ctxOKTry} = \reachable{\textsc{[err]}}$ and $\textit{ctxTry} = \ctxPositions{{E}}{try}{\{1\}}$. 
%
\[
%\magenta{
\ninference{ctx-compliant}
{ 
P = \textit{ctxTry} \land \textit{tryF}\, \\ {\ctxPositions{\userLan{E}}{try}{\{1\}}} \in P \\\\
er \grammarDerivationG\key{error} \myimplies 1 \in \{1\} \textit{ (* \key{error} is at position 1 *)}
}
{\assert
%{\ctxPositions{\userLan{E}}{app}{\{1,2\}}}
{P}
{\textsc{[err]}:\try{\key{error}}{e}, s  \step  e, s}
{P\land \reachable{\textsc{[err]}}}}
%}
\]

\section{Example of a Proof Rule with Negated Assertions}\label{negation}

%As in program logics, the proof rule \textsc{(consequence)} makes use of implication $ \Rightarrow$. 
As in program logics, \textsc{(consequence)} uses implication $ \Rightarrow$. 
%As in program logics, \textsc{(consequence)} makes use of implication $ \Rightarrow$. 
The first premise $P \Rightarrow P'$ of \textsc{(consequence)} can be the implication 
%The first premise $P \Rightarrow P'$ of \textsc{(consequence)} can be used for the implication 
%The first premise of \textsc{(consequence)}, i.e., $P \Rightarrow P'$, can be used for the implication 
%The first premise of proof rule \textsc{(consequence)}, i.e., $P \Rightarrow P'$, can be used for the implication 
$\ctxPositions{\userLan{F}}{try}{\{\}} 
\Rightarrow
\lnot\ctxPositions{\userLan{F}}{try}{\{1,2\}}$. 
%
%This implication can be read as 
This implication holds, as it says 
\quoting{since 
%\quoting{Since 
no arguments of \textit{try} is $F$, it is not the case that the arguments at positions $1$ and $2$ are $F$.} 
Proof rule \textsc{(error-handler)} could be equivalently written in the following way. (Notice $i \in I$ rather than $i \not\in I$.)

\[
\ninference{error-handler-with-negation}
{ \reachable{rn} \in P \\
(\lnot\ctxPositions{\userLan{F}}{op}{I}) \in P \\\\
err \grammarDerivationG t_i 
%\\\\
\\
%I_{err} \subseteq I_{ctx} \\ 
i \in I
}
{\assert{P}{rn: (op \app t_1, \ldots, t_n), \widetilde{s} \step t,  \widetilde{s'}}{P \land \errHandler{op}{i}}}
\]

This was our first design of \textsc{(error-handler)} but we ultimately decided that it was unnecessarily complicated to require a step through \textsc{(consequence)}, 
and we simplified the proof rule. 
For the sake of generality, we have kept negated assertions $\lnot P$ in the grammar. % of assertions. 

\section{Example of Language Definition in $\lna$}\label{toolLanguage}

%Below, we show some examples of algorithmic typing rules (other than those of $\runningExample$) that we have used in our tested language definitions. 
%Our extension of $\lnp$ uses \texttt{|a-} to denote the algorithmic typing relation and \texttt{<a} to denote the algorithmic subtyping relation. 

Language definitions are text files with extension \texttt{.lan}. 
The following is the language definition for the $\lambda$-calculus (with booleans as base case). 
The \quoting{\texttt{<==}} symbol should be read as \quoting{provided that}. 
Rule \textsc{[t-var]} is automatically considered. % in the language and so language designers do not need to write it. 
%Rule \textsc{[t-var]} is automatically considered in the language and so language designers do not need to write it. 
%
%\begingroup
%    \fontsize{9.5pt}{9.5pt}\selectfont
%\scalebox{0.9}{
%\resizebox{0.85\textwidth}{!}{
%\newlength\someheight
%\setlength\someheight{5cm}
%\begin{adjustbox}{width=0.9\textwidth,height=\someheight,keepaspectratio}
%\hspace*{+30cm}
{\small
\begin{lstlisting}[xleftmargin=-0.1cm,keywordstyle=\color{black},xrightmargin=-1cm,basicstyle=\ttfamily]
Expression E ::=  true | false | x | (abs T (x)E) | (app E E)
Type T ::= bool | (arrow T T)
Value V ::=  true | false | (abs T (x)E)
Context C ::= [] | (app C E) | (app V C)

[T-TRUE]
Gamma |- true : bool.

[T-FALSE]
Gamma |- false : bool.

[T-ABS]
Gm |- (abs T (x)E) : (arrow T T') <== Gm, x : T |- E : T'.

[T-APP]
Gamma |- (app E1 E2) : T2 <== Gamma |- E1 : (arrow T1 T2) 
                              /\ Gamma |- E2 : T1.
[BETA]
(app (abs T (x)E) V) --> E[V/x] <== value V.
\end{lstlisting}
}
%\end{adjustbox}
%}
%\endgroup

\noindent (This type of textual form for languages is not a novelty of $\lna$. It is indeed inspired by the style that the Ott tool\footnote{https://github.com/ott-lang/ott} has adopted long ago.)

\end{document}